\documentclass[12pt]{JHEP3}

\usepackage{amsmath,amssymb,epsf,graphicx}

\newcommand{\bsp}{\begin{split}}
\newcommand{\esp}{\end{split}}
\def\one{{\,\hbox{1\kern-.8mm l}}}
\newcommand{\Dslash}{\not{\hbox{\kern-4pt $D$}}}
\newcommand{\pdslash}{\not{\hbox{\kern-2pt $\partial$}}}
\newcommand{\cL}{\mathcal{L}}

\newcommand{\Comment}[1]{{}}

\def\IZ{{\mathbb Z}}

\def\IR{{\mathbb R}}

\newcommand{\bc}{\begin{center}}
\newcommand{\ec}{\end{center}}
\newcommand{\ba}{\begin{array}}
\newcommand{\ea}{\end{array}}
\newcommand{\beq}{\begin{equation}}
\newcommand{\eeq}{\end{equation}}
\newcommand{\bea}{\begin{eqnarray}}
\newcommand{\eea}{\end{eqnarray}}
\newcommand{\bmx}{\begin{pmatrix}}
\newcommand{\emx}{\end{pmatrix}}

\newcommand{\be}{\begin{equation}}
\newcommand{\ee}{\end{equation}}

\newcommand{\del}{\partial}
\newcommand{\half}{{\frac{1}{2}\,}}
\newcommand{\tr}{{\rm tr}}

\newcommand{\eref}[1]{Eq.\,(\ref{#1})}

\def\IB{\relax{\rm I\kern-.18em B}}
\def\IC{{\relax\hbox{\kern.3em{\cmss I}$\kern-.4em{\rm C}$}}}
\def\ID{\relax{\rm I\kern-.18em D}}
\def\IE{\relax{\rm I\kern-.18em E}}
\def\IF{\relax{\rm I\kern-.18em F}}
\def\II{\relax{\rm I\kern-.18em I}}
\def\IZ{\relax{\sf Z\kern-.35em Z}}
\def\Id{\relax{1\kern-.32em 1}}
\def\IG{\relax\hbox{$\inbar\kern-.3em{\rm G}$}}
\def\IR{\relax{\rm I\kern-.18em R}}
\newcommand\sfrac[2]{{\textstyle\frac{#1}{#2}}}

\newcommand\shalf{{\textstyle\frac12}}

\normalsize

\font\litfont=cmmi7 scaled\magstephalf

\title{\bf String theory: a perspective over 
the last 25 years}

\author{{\bf Sunil Mukhi}\\
{\em Tata Institute of Fundamental Research, Mumbai}}

\date{} \abstract{This article provides some historical background and
  then reviews developments in string theory over the last twenty-five
  years or so. Both perturbative and non-perturbative approaches to
  string theory are surveyed and their impact on how we view quantum
  gravity is analysed.\footnote{Invited review, 
published in Classical and Quantum Gravity, {\bf 28} (2011) 153001.}} 

\preprint{TIFR/TH/11-44}

\begin{document}

%\advance\baselineskip by 3pt

%\maketitle

\section{Origins}

%%%%%%%%%%%%%%%%%%%%%%%%%%%%%%%%%%%%%%%%%%%%%%

String theory was conceived partly by accident and partly by design.
It arose in the study of scattering amplitudes for hadronic bound
states.  Lacking a fundamental formulation (that today is believed to
be Quantum Chromodynamics), hadronic amplitudes were studied in the
1960's using symmetry, consistency and some observed properties such
as the asymptotic rate of growth of cross-sections with energy.  A
beautiful formula proposed in
1968\cite{Ademollo:1969dd,Veneziano:1968yb} embodied the expected
properties of 4-point amplitudes. The Veneziano formula eventually
turned out to describe the scattering of open relativistic
strings. These were later generalised\cite{Virasoro:1969me} in a way
that is now understood to describe closed-string
scattering\footnote{Details on the early history of string
  theory, along with references, can be found in
  Ref.\cite{Schwarz:2007yc}.}.

The proposed hadronic amplitudes depended on the familiar Mandelstam
variables \be s\equiv (k_1+k_2)^2,\qquad t\equiv (k_1-k_3)^2,\qquad
u\equiv (k_1-k_4)^2 \ee where $k_i$ are the momenta of the two
incoming and two outgoing particles, and also on a parameter $\alpha'$
with dimensions of ${\rm (length)}^2$. $\alpha'$ is phenomenologically
defined as the slope of ``Regge trajectories'', approximately linear
plots of the spin of different hadronic resonances
versus their ${\rm (mass)}^2$. A basic reason why Regge trajectories
should be linear was not known, nor was it clear what $\alpha'$ means
in fundamental terms.

Over the next couple of years, Susskind\cite{Susskind:1970xm} and
independently Nambu\cite{Nambu:1970si} proposed a way to understand
the Veneziano formula, and thereby the strong interactions, from a
novel starting point. They postulated the existence of a
``fundamental'' relativistic string to represent the confining flux
between quarks. Quantising this string would lead to physical
excitations that could be identified with the baryonic and mesonic
states formed by bound quarks.

The world-sheet action\footnote{For derivations and explanations of
  the results in this section, with references, see for example
Ref.\cite{Green:1987sp}.}
describing a free string propagating in flat spacetime can be written:
\be 
S=\int d\sigma dt\, \sqrt{-\det (\del_a X^\mu\,\del_b X^\mu)} 
\ee
where $(\sigma,t)$ are the parameters labelling points on the string
world-sheet, $X^\mu$ are the space-time coordinates of the string and 
$\del_a=(\del_\sigma,\del_t)$. The integral is the invariant
area of the world-sheet in the pull-back of the flat Minkowski metric
of space-time. The action possesses reparametrisation invariance on
the world-sheet and by a suitable gauge-fixing of this invariance it
can be brought to the simpler form: \be S=\shalf\int d\sigma dt~
\del_a X^\mu\, \del^a X_\mu \ee subject to constraints. This action is
conformally invariant at the classical level.

The first success of string theory was in explaining Regge behaviour.
This follows from the fact that the oscillator modes
$a^{\dagger\,\mu}_n$ that create excitations of the string, being
Fourier modes of the string coordinate $X^\mu(\sigma,t)$, carry a
spacetime index $\mu$. Applying a particular oscillator $n$ times to
the ground state $|0\rangle$ creates a state: \be
|\mu_1\mu_2\cdots\mu_p\rangle=
a^{\dagger\,\mu_1}_{n}a^{\dagger\,\mu_2}_{n}\cdots
a^{\dagger\,\mu_p}_{n}|0\rangle \ee with $n$ symmetrised spacetime
indices. This means that the spin of the state is $n$. At the same
time the ${\rm (mass)}^2$ operator for a string is the number operator
for these oscillators in units of the string tension. Hence the above
state has ${\rm (mass)}^2\sim nT$ with $T$ being the string tension.
Regge behaviour then follows immediately and we see that the Regge
slope is $\alpha'=1/T$. Note that
in units where $\hbar=c=1$, tension indeed has dimensions of $({\rm
  length})^{-2}$ and hence $\alpha'$ has dimensions of ${\rm
  (length)}^2$. It is a fundamental constant of string theory. 

Quantising the fundamental string carefully using different
methods confirmed the expectations: strings display Regge behaviour
and the scattering amplitudes of low-lying strings states are, at tree
level, those postulated by Veneziano and Virasoro along with their
generalisations. 

Some surprises also emerged. At the quantum level, conformal
invariance of the string world-sheet action turns out to have a
quantum anomaly that vanishes only in a spacetime of 26 dimensions. In
the presence of the anomaly, standard quantisation of the string is
inconsistent (a mode describing scale fluctuations of the world-sheet
fails to decouple). Therefore string theory is, in particular,
inconsistent in four space-time dimensions.

There is also another problem. Even in 26 dimensions, the particle
spectrum of the string starts with a state of tachyonic (imaginary)
mass. This arises from the combined effect of the zero-point energies
of infinitely many oscillators describing the independent modes of the
string. Tachyons are present in the spectrum of both open strings,
which are interpreted as carrying quarks at their ends, and closed
strings, which describe the quark-less sector of the theory of strong
interactions -- what we would today call the glueball spectrum.

A tachyon in the particle spectrum of a theory does not necessarily
mean the theory is inconsistent. This is familiar in ordinary quantum
field theory, where a mass term of the ``wrong'' sign (equivalent to
making the mass imaginary) is interpreted as an instability of the
theory {\em when expanded above a vanishing expectation value of the
  field}.  Such theories are generally consistent when we expand the
field about its true (non-zero) vacuum expectation value.  This
phenomenon is well-understood in quantum field theory, and is the
basis of the Higgs mechanism crucial to the Standard Model of particle
physics. Unfortunately it is much harder to understand explicitly in
the context of string theory.  Shifting a field by its own expectation
value in string theory requires knowledge of the ``off-shell''
dynamics of strings, which was certainly not known at the time string
theory was first studied. In the last decade or so, however, more
powerful techniques\cite{Sen:1999nx,Rastelli:2001uv} have taught us
that the tachyonic string theory described above (now known as the
``bosonic string'') is indeed inconsistent, because the analogue of
the Higgs potential is unbounded below and the expectation value of
the field therefore ``runs away'' to infinity.

Analysis of the spectrum of open and closed strings reveals that while
both contain tachyons, both also contain conventional massless and
massive particles. In particular open strings produce a spin-1
massless particle while closed strings produce a massless particle of
spin 2 (spin is of course measured in units of $\hbar$). Such
particles do not have any obvious counterpart in the hadron spectrum.
As we will see below, this fact led to a radical new role for string
theory: as a theory of fundamental processes.

The consistency requirements of 26 dimensions as well as the presence
of a tachyon and a massless spin-2 particle proved rather discouraging
for the string approach to hadronic physics. The focus of strong
interactions therefore soon shifted, with the quark model slowly
becoming a reality and being described successfully by a gauge field
theory, quantum chromodynamics. In this period the string idea of
Susskind and Nambu remained a useful qualitative picture to model
the force between quarks. The fact that under normal conditions quarks
are permanently confined fits nicely with the notion that they are
connected by strings of fixed tension, because then increasing
the inter-quark separation gives rise to a linearly growing potential
energy which is naturally expected to confine.

\section{Superstrings}

\subsection{The fermionic string}

A small group of physicists continued to think about strings in the
1970's\cite{Schwarz:2007yc} 
although by then the focus of particle physics had shifted to
gauge theory. They investigated the formalism of string
theory with a view to addressing its negative features, primarily the
tachyon and the requirement of 26 dimensions. Also the original string
theory did not contain fermions in its spectrum, so the inclusion of
fermions became one of the goals.

It was a natural step to introduce fermionic degrees of freedom on the
string world-sheet. These were to be thought of as ``fermionic
coordinates'' $\psi(\sigma,t)$ partnering the usual bosonic string
coordinates $X^\mu(\sigma,t)$. A proper treatment of the fermionic
string took some time to evolve but by the end it was clear that some
of the excitations of this string were indeed fermions in space-time.
Many remarkable features emerged from this somewhat formal exercise.
The world-sheet theory of the fermionic string exhibited an entirely
novel symmetry (at the time) called ``supersymmetry'', that related
bosons to fermions. Moreover the particles that arose by quantising
this string were also related to each other by supersymmetry, but now
in space-time. The fermionic degrees of freedom modified the
condition that had required 26 space-time dimensions in the original
string theory. The required number of dimensions was now 10
-- still far from the real-world value of 4, but considerably closer.

Space-time supersymmetry has an immediate and striking consequence.
In supersymmetric systems the contributions of bosonic and fermionic
degrees of freedom to the zero-point energy cancel each other out.
This result immediately implies the absence of tachyons in the new
``superstring'' theory. Thus three distinct problems: a very high
critical dimension of 26, the presence of tachyons and the absence of
fermions, were all simultaneously solved by the advent of
superstrings.

\subsection{Gravity and gauge symmetry from superstrings}

Given that the original motivation for string theory had been to
explain the forces between quarks and the formation of hadrons, the
advent of superstring theory seemed to take the subject in an entirely
new direction. It could not be applied to the strong interactions for
several ``obvious'' reasons: it was still not consistent in 4
spacetime dimensions, and it had spacetime supersymmetry which is not
a property of the strong interactions at least at low energies. But
the most undesirable property was that, upon quantisation, the free
closed superstring had a massless spin-2 excitation, while the free
open superstring had a massless spin-1 excitation. These were features
of the original bosonic string theory that, unlike the tachyon, did
not go away by introducing supersymmetry, and they seemed to have
little to do with hadrons.

In quantum field theory, fields of integer spin very generally have a
problem with negative-norm states. The reason is that the norm of a
state carrying vector indices can only be Lorentz invariant if it is
expressed in terms of the Minkowski metric. For example a vector state
$|\mu\rangle$ or a tensor state $|\mu\nu\rangle$ will have norms
proportional to:
\be
\bsp
\langle\mu|\nu\rangle &\sim\eta_{\mu\nu}\\
\langle\mu\nu|\rho\sigma\rangle &\sim\eta_{\mu\rho}\eta_{\nu\sigma}+\cdots
\end{split}
\ee where in the second line there can be extra terms to reflect the
symmetry/antisymmetry of the state if any.  Because $\eta_{\mu\nu}$
has a negative component, the above formula always leads to a
negative-norm state. As a consequence, unitarity of the theory
becomes problematic.

The solution to this problem is that fields of integer spin must have
a local invariance. This can project out the negative-norm states and
restore unitarity. In the case of spin-2, the local invariance is that
of general coordinate transformations and the resulting theory is
general relativity. With this symmetry, unphysical modes of a spin-2
particle that would have had null or negative norm decouple and the
particle is interpreted as a graviton.  This line of thought motivated
the suggestion that a similar mechanism may hold for superstrings. If
so, one could interpret the spin-2 particle in the closed string
spectrum as a graviton. Closed string theories would then be theories
of gravity. By the same token, since open strings have massless spin-1
particles in their spectrum and these are consistent only in the
presence of a local gauge symmetry, one would expect that open strings
describe gauge particles.

These rather audacious proposals could be subjected to very stringent
tests. If closed strings described gravity and open strings described
gauge interactions, it must be that the string-string interactions
contain a tightly correlated set of terms dictated by the symmetries
of these interactions. For example, general coordinate invariance
predicts a unique interaction vertex among four gravitons at
two-derivative level (i.e.  having two factors of momentum in the
interaction). This is unique both in its coefficient (once the
propagator and three-point function are normalised) and in its
tensorial structure. 

Now, unlike in quantum field theory where one is free to add any
interaction one likes (at least at tree level), in string theory there
is a unique prescription\cite{Green:1987sp,Polchinski:1998rq} to
compute scattering amplitudes for any fixed background.  For
four-point amplitudes at tree level, this amounts to expanding the
superstring analogue of the Virasoro amplitude and keeping the leading
term in the parameter $\alpha'$. From this one reads off the
interaction term in an effective Lagrangian for the corresponding
massless particle. Remarkably this turns out to have exactly the
structure predicted by general relativity.

The expansion of the Einstein-Hilbert Lagrangian 
\be 
\sqrt{|g|}R 
\label{einhil}
\ee in powers of the fluctuation field $h_{\mu\nu}$ defined by: \be
g_{\mu\nu}=\eta_{\mu\nu}+h_{\mu\nu} \ee is well-known to contain
infinitely many powers $(h_{\mu\nu})^n$, with definite index
contractions, corresponding to contact interactions of $n$ gravitons
for every $n$. Therefore to establish rigorously that general
relativity is reproduced by the tree-level scattering of closed
strings, one would need to calculate $n$-point closed-string
amplitudes for all $n$, a rather difficult task. Nevertheless every
term that has been computed leads, after performing allowed field
redefinitions, to the correct expression in the expanded
Einstein-Hilbert action. So there can no longer be any doubt that
closed superstrings describe gravity\footnote{For bosonic strings, 
  on ignoring the tachyon one finds that they also describe gravity in
  more or less the same way. Now the tachyon can be meaningfully
  ignored at tree level, but once quantum (loop) corrections are
  included then the tachyon renders the theory inconsistent. Hence we
  confine our main remarks to the tachyon-free superstring.}.

It is not just Einstein-Hilbert gravity that is described by strings,
however. The scattering amplitudes for closed strings have
an expansion in powers of $\alpha'$. Because $\alpha'$ has
dimensions of $({\rm length})^2$, these terms must have additional
powers of momenta, or in position space, {\em derivatives}. These
higher-derivative terms too are found to be
general-coordinate-invariant. Since they can be neglected for
sufficiently slowly varying fields, the correct statement is that
closed string amplitudes reproduce Einstein-Hilbert gravity for slowly
varying fields, with calculable corrections expressed in terms
higher-order variations of the fields.

A very similar calculation for open strings reveals that to lowest
order in $\alpha'$ one finds a non-Abelian gauge theory of
Yang-Mills type, with its characteristic cubic and quartic
self-interactions. Again there are higher derivative corrections that
are suppressed for slowly varying fields. In the simplest (Abelian)
case, an infinite subset of these turns out to be in correspondence
with the old non-linear electrodynamics theory of Dirac and of Born and
Infeld\cite{Polchinski:1998rq}. 

The more general non-Abelian calculation was initially performed using
a somewhat ad hoc prescription, introducing matrices called ``Chan-Paton
factors'' at the ends of the open string. In the presence of such
matrices, one automatically obtained a $U(N)$ gauge group for $N\times
N$ matrices.  The introduction of unoriented open strings enabled one
to realise orthogonal and symplectic groups as well. Much later it
became clear that these Chan-Paton matrices were labelling dynamical
objects on which open strings end, or ``D-branes'', about which we
will say more below.

Thus the proposal that closed and open strings describe gravity and
gauge theory was convincingly supported by amplitude calculations.
Because all known fundamental interactions fall into these
two types, it is no surprise that string theory emerged as a natural
framework to describe the world.

\subsection{Dilaton and the genus expansion}

So far we have spoken only about tree amplitudes. In field theory,
``tree level'' refers to the lowest order in an expansion in a coupling
constant. In string theory, by contrast, there is no parameter in the
theory other than the dimensional constant $\alpha'$. Therefore ``tree
level'' as discussed above refers simply to a computation where the
world-sheet has a tree-like configuration without any holes or
handles. For a beautiful reason, this in fact turns out to
be the lowest order in a ``hidden'' string coupling.

This arises as follows. Along with a graviton, quantisation of the
closed string gives rise to a massless scalar particle called the
dilaton\cite{Green:1987sp,Polchinski:1998rq} denoted $\Phi$. In
perturbation theory it turns out that this scalar has no potential,
and is therefore the first of many scalars in string theory (called
``moduli'') that can take arbitrary vacuum expectation values.

From the unique structure of string interactions one can infer that in
tree diagrams the dilaton couples to the low-energy Lagrangian by an
overall multiplicative factor of $e^{-2\Phi}$. The vacuum expectation
value $\langle\Phi\rangle=\Phi_0$ can be taken outside action and gets
identified with $\frac{1}{g^2}$ where $g$ is a coupling constant,
since that factor (in some frame) is what multiplies the whole action
of a field theory. Thus by virtue of the dilaton VEV, string theory
acquires a coupling constant:
\be
g=e^{\Phi_0}
\ee

Now if we compute scattering amplitudes on world-sheets of some higher
genus $h>0$, it can be shown that the power of the dilaton VEV multiplying
the action is $e^{(2h-2)\Phi_0}$. It follows that world-sheets of
genus $h$ are weighted by $g^{2h-2}$ and therefore the perturbation
expansion of string theory is an expansion in world-sheets of
increasing genus:
\bc
\includegraphics[height=2.4cm]{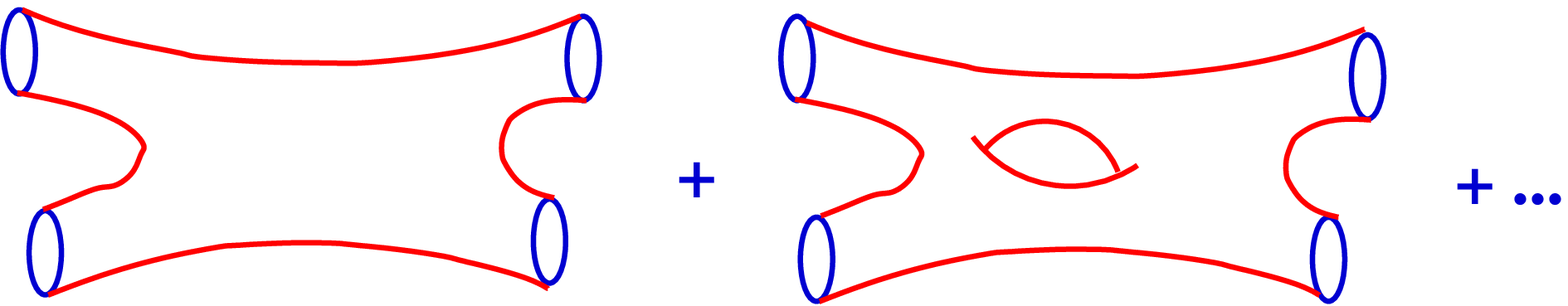}\\[2mm]
Order 1
~~~~~~~~~~~~~~~~~~~~~~~~~~~~~~~~ Order
$(g^2)$~~~~~~~~~
\ec
In each order one has to sum over all surfaces of the given
topology. This sum is subtle for various mathematical and physical
reasons, but has by now been thoroughly investigated.

\subsection{Quantum corrections}

Once it was clear that superstring theory is consistent at tree level,
it became extremely urgent to check for its consistency after
introducing perturbative quantum corrections, known in field theory as
``loop'' corrections\cite{Green:1987mn,Polchinski:1998rr}. The rules
for tree level superstring amplitudes are particularly easy to extend
to the next order in perturbation theory, namely one-loop level
corresponding to genus-one Riemann surfaces. Here the proposal that
closed strings describe gravity can receive its first test at the
quantum level.

Einstein-Hilbert gravity suffers from ultraviolet divergences in every
order of perturbation theory. From this it follows that at least in a
perturbation expansion the theory is non-predictive.  String theory
could well turn out to have the same problem. However, there are
physical grounds for optimism. One may hope that the spatial extent of
the string (parametrised by $\alpha'$) could provide an ultraviolet
cutoff. Then graviton scattering in superstring theory would have
finite loop corrections. This hope was dramatically realised with the
computation of one-loop graviton amplitudes in superstring theory,
which indeed turned out to be ultraviolet finite.

In field theory, it is known that combining supersymmetry with gravity
leads to the theory of ``supergravity'' in which gauged or local
supersymmetry is present. Therefore just on symmetry grounds, one can
show that the low-energy effective action of superstring theories must
correspond to a field theory of supergravity, possibly coupled to
supersymmetric matter (and augmented by higher-derivative
corrections). Now, important phenomenological work using supergravity
theories was done in the late 1970's and early 1980's(see for example
Ref.\cite{Nath:1983fp}), ignoring the fact that -- as field theories
-- they were probably not ultraviolet complete. In a remarkable
convergence of streams of research, superstrings provided a sound
justification for this work: the interesting physical applications of
supergravity at low energy could co-exist with an ultraviolet-finite
high-energy completion coming from string theory.

Thus superstring theory was understood to be, in 10 dimensions, a
quantum theory of gravity incorporating non-Abelian interactions and
supersymmetry in a unified framework.  With gravity and non-Abelian
gauge symmetry being undisputed properties of nature, and
supersymmetry also a likely property, superstring theory exhibited the
potential to answer all the important questions about fundamental
interactions that had remained unsolved for over half a century.
Moreover it was a theoretical formulation of unmatched beauty and
power. However it was very far away from being able to reproduce even
gross properties of the real world such as the observed gauge groups
and parity-violating particle spectrum.

There were few people working on superstrings at the time this state
of affairs was reached. Particle theorists were more excited, and with
good reason, over the recent successes of the electro-weak theory as
well as quantum chromodynamics. Moreover, the calculational techniques
of string theory as understood at the time were unfamiliar and
difficult to grasp. And of course gravity was totally irrelevant to
the particle-physics experiments which these theories could explain so
convincingly.

\subsection{Conformal invariance and the sigma-model approach}

Stepping briefly out of historical sequence, we mention here a
slightly different approach to studying strings that provides more
direct evidence of the presence of gravity and general coordinate
invariance. In this approach, instead of studying strings propagating
in flat spacetime one writes down the world-sheet action for strings
propagating in an arbitrary curved space-time, and more generally in
arbitrary background fields.  In the specific case of a nontrivial
space-time with metric $g_{\mu\nu}(X)$ (in some coordinate system),
the world-sheet theory reduces to a non-linear sigma-model in
two dimensions, with Lagrangian: \be {\cal L}=\half
g_{\mu\nu}(X)\del_a X^\mu \del^a X^\nu+\cdots \ee where as before
$X^\mu(\sigma,t)$ is the position coordinate of the string and the
world-sheet derivative is $\del_a=(\del_t,\del_\sigma)$. The terms
represented by $+\cdots$ are those which depend on the fermionic
coordinates of the superstring.

When $g_{\mu\nu}(X)$ is a non-trivial function of $X$, the above
action describes an interacting two-dimensional field theory in which
the coordinates $X^\mu$ play the role of scalar fields. Because it has
no dimensional coupling\footnote{This is potentially confusing because
  the coupling is really $\sqrt{\alpha'}$ which multiplies each
  occurrence of $X^\mu$. The explanation is that from the space-time
  point of view both $\alpha'$ and $X^\mu$ are dimensional, with
  $\sqrt{\alpha'}X^\mu$ being dimensionless. But from the world-sheet
  point of view both $\alpha'$ and $X$ are classically dimensionless,
  this being true for $X^\mu$ by virtue of its identification with a
  scalar field in 2 dimensions.}, this theory is classically conformal
invariant just like the one for strings propagating in flat space-time
that we discussed earlier.  We already saw that conformal invariance
is generically violated and renders the theory inconsistent unless the
dimension is critical. In the present case there are additional
sources of anomalies, so not only does anomaly-freedom fix the
critical dimension to a critical value (10 for superstrings) but it
also imposes {\em conditions on the allowed metric $g_{\mu\nu}(X)$}. Indeed,
to first order in $\alpha'$, a short calculation\cite{Friedan:1980jf,
  AlvarezGaume:1981hn} reveals that the theory remains conformally
invariant if: 
\be 
R_{\mu\nu}=0 
\ee 
Notice that this is Einstein's equation in a vacuum! This resemblance
is no coincidence. It has been convincingly
argued\cite{Sen:1985qt,Callan:1985ia} that the condition for conformal
invariance imposed on the string sigma-model is equivalent to the
equations of motion of the effective low-energy field theory arising
from the string theory. In particular for backgrounds with non-trivial
stress-energy, the conformal invariance condition becomes the standard
Einstein equation with the stress-energy on the right hand side: \be
R_{\mu\nu}= 8\pi G_N\Big(T_{\mu\nu}+\sfrac{1}{2-D}\,g_{\mu\nu}T\Big)
\ee

Thus, conformal invariance of the string world-sheet provides a new
principle, without an analogue in particle mechanics, to derive
space-time actions. In particular it shows that the low-energy
effective action of closed string theory, to lowest order in
$\alpha'$, is given by the Einstein-Hilbert action \eref{einhil}. This
term is universal for all superstring (and even bosonic string)
theories. But from the next order in $\alpha'$ there are differences
depending on whether one does or does not have supersymmetry. Some
corrections that arise in the bosonic string do not arise for the
superstring\cite{AlvarezGaume:1981hn,Gross:1986iv}.  This has crucial
consequences for the physics of the theory: in the superstring,
space-times that solve the equations of motion to lowest order in
$\alpha'$ can continue to solve them when higher-order corrections are
included. This is not so in the bosonic string, where higher-order
corrections rule out a large class of potentially interesting
solutions.

\subsection{New insights about gravity: I}

The developments reviewed above gave rise to several important
insights into the nature of quantum gravity. Many were properties of
gravity that had already been suspected or conjectured but for which
string theory provided a concrete realisation and/or tangible new
evidence.

(i) It is possible to have a well-defined ultraviolet finite theory of
gravity, with the characteristic minimum size providing a UV cutoff.
Even if the idea is not totally new, string theory provides a precise
mechanism and allows one to dissect precisely how the cutoff works.
The world-sheet of the string is a Riemann surface on which certain
diffeomorphisms called ``modular transformations'' act. While summing
over all surfaces, the dangerous ones for UV divergences are those
that become very ``thin'', corresponding to high momentum flowing
through. But a modular transformation relates such ``thin'' surfaces
to others that are not degenerating, so we never actually reach
infinitely thin surfaces, or equivalently infinitely high momenta.

(ii) Higher derivative terms are generic.  Any effective action will
have this property, but in string theory one sees for the first time a
concrete computation of these terms given a particular space-time
background.

(iii) Space-time is an option. Classical solutions are configurations
that give a conformal-invariant world-sheet, and these need not
resemble space-time (for example, a tensor product of ``minimal''
conformal field theories with total central charge equal to the
critical value but each component having central charge less than 1
would be a valid solution).

(iv) Gravity and gauge symmetry have a common origin. With closed and
open strings, the identical quantisation procedure yields respectively
gravitons and gauge bosons. Moreover gravitons arise by combining a
``vector'' state from the left- and right- movers of the closed string,
making them in some sense the ``square'' of gauge fields. Present-day
research on gluon and graviton amplitudes (see for example 
Refs.\cite{Bern:2002kj,Ananth:2010uy} has revealed more and more
structures suggesting this ``square'' relationship between gravity and
gauge theory. A deep understanding of this could revolutionise our
understanding of the gravitational force.

\section{Structure and varieties of superstring theories}

\subsection{Five types of superstrings}

String theories, like field theories, have supersymmetry generators
that are organised as spinors of the relevant local Lorentz group.
Thus in 10 dimensions, a supersymmetry generator (or ``supercharge'')
$Q_\alpha$ is a real 16-component Majorana-Weyl spinor of the local
Lorentz group $SO(9,1)$. Now it was found (for details, see for
example Refs.\cite{Green:1987sp,Polchinski:1998rr}) that string theory
could accommodate either one or two such spinor generators as a
symmetry\footnote{The same is true in 10d field theory.}, 
leading to type I and type II superstring theories respectively.

It was also shown that type II theories can only have closed strings,
because open-string boundary conditions break supersymmetry\footnote{Much
later it was understood that there are actually {\em sectors} in type II
theories that admit open strings. The physical interpretation of these
sectors is in terms of dynamical objects called D-branes that we will
discuss below.}. Moreover, the two Majorana-Weyl supersymmetries can 
correspond to a pair of spinors of opposite chirality or the same
chirality, leading to two physically inequivalent theories called type IIA
and type IIB. The latter is parity-violating, since both supercharges
have the same chirality.

In type I theories one has open strings which, via the Chan-Paton
mechanism discussed above, are associated to a gauge group\footnote{In
  a modern interpretation the presence of open strings is interpreted
  in terms of ``condensed D-branes in the vacuum''.}. The Chan-Paton
mechanism is able to accommodate only gauge groups of the unitary,
orthogonal or symplectic types. Thus, at least classically, one can
have infinitely many type I superstring theories in 10 dimensions, one
for each choice of gauge group. Having one chiral supercharge, they
are all parity-violating in 10 dimensions.

An exotic way to achieve a ``type I-like'' theory is to match
the left-moving sector of the type II superstring to the right-moving
sector of the bosonic string. This leads to the so-called
``heterotic'' superstring theory\cite{Gross:1984dd}. 
Because it uses only half the type
II superstring, in a precise technical sense, it has half the
supersymmetries and therefore has ${\cal N}=1$ supersymmetry in 10
dimensions. For historical reasons heterotic strings are not called
``type I'' because the latter terminology is reserved for string
theories having both open and closed string excitations about the
vacuum. 

As one would expect, the low-energy limits of type I/II string theories
are type I/II supergravity field theories in 10 dimensions. The field
theories contain the massless fields associated to the string. Because
supersymmetry is so tightly constraining, one can actually write out
the entire low-energy action for type I/II strings at the classical
level and to lowest order in $\alpha'$ purely on  grounds of
supersymmetry. 

Quantum mechanically, it is essential to check that the
parity-violating string theories are free of anomalies. Because
anomalies do not depend on the ultraviolet behaviour of a theory, the
anomaly structure is the same for superstring theories and for their
low-energy supergravity field theories. A search was therefore
launched in the early days for possible anomaly-free supergravity
theories (possibly coupled to supersymmetric Yang-Mills theories) in
10 dimensions. Type IIA is trivially anomaly-free, while type IIB
supergravity theory was found\cite{AlvarezGaume:1983ig} to be
anomaly-free due to non-trivial cancellations among a number of
anomalous contributions.

Later Green and Schwarz\cite{Green:1984sg} studied the type I case.
They made the remarkable discovery that there were precisely two
anomaly-free field theories in 10 dimensions, one with gauge group
$SO(32)$ and the other with the exceptional gauge group $E_8\times
E_8$. On trying to match these with string theories they found that an
open-and-closed type I theory could be constructed to have gauge group
$SO(32)$. However, no such theory with Chan-Paton factors can have the
gauge group $E_8\times E_8$. Heterotic strings, however, can
incorporate either of the two gauge groups. Thus at the end we have
five superstring theories in 10 dimensions: type IIA, IIB, type I
($SO(32)$), heterotic ($SO(32)$), heterotic ($E_8\times E_8$).

The ``almost uniqueness'' of superstring theory, including a tight
restriction on gauge groups in the type I case, was very exciting.  A
selection principle governing the gauge groups that can occur in
nature would be quite unprecedented. Such a principle has never
existed in four dimensional field theory, where the constraints of
anomaly freedom are still present but they restrict the
representations that can appear rather than the groups themselves.

The elegant classification into five distinct superstring theories was
somewhat tempered by the later discovery of non-supersymmetric strings
that, like superstrings, are tachyon-free and whose critical dimension
is still 10. These theories were found by extracting different
space-time dynamics from a common world-sheet theory by imposing
different projections on it.  These discoveries presaged the advent of
duality, which made it very clear that essentially all string theories
are different vacua of a common theory.

\subsection{Tensors and Ramond-Ramond fields}

In addition to the graviton, the ground state of the closed string
contains a whole supermultiplet of massless fields in 10 dimensions.
With maximal or ${\cal N}=2$ supersymmetry, this supermultiplet
includes the graviton, the dilaton and several tensor fields of the
type $C_{\mu_1\mu_2\cdots \mu_p}$ that are totally antisymmetric in
their indices (see for example Ref.\cite{Polchinski:1998rr}). These
are referred to as ``$p$-forms''. Some of these $p$-forms are special
in the way that they arise from string quantisation and the nature of
their associated gauge invariance (a point that is too technical to
discuss here). They carry the name ``Ramond-Ramond (RR) fields''. Type
IIA strings have RR 1-form and 3-form fields while type IIB strings
have RR 0-form (scalar), 2-form and 4-form fields.

To illustrate the structure of the low-energy theory, we quote
here the action of type IIA supergravity keeping only the RR 1-form
and ignoring the other tensor fields and fermions:
\be S_{\hbox{\litfont
    type}\,\hbox{\litfont IIA}}~=~ {1\over (2\pi)^7 \ell_s^8} \int
d^{10} x\sqrt{-\|G\|}\Bigg[~e^{-2\Phi} \left(R + | d\Phi|^2\right) -
{2\over 8!}| dA|^2\Bigg] 
\ee
where we have used the notation $\ell_s =\sqrt{\alpha'}$.  Here
$G_{\mu\nu},\Phi$ are the graviton and dilaton while $A_\mu$ is the RR
1-form. We will return to this action in a subsequent section.

\subsection{Perturbative dualities}

String amplitudes possess the attribute of ``channel duality'' which
can be summarised as follows\cite{Schwarz:2007yc,Green:1987sp}. In
particle physics, two-particle to two-particle scattering takes place
in one of three ways, corresponding to the $s,t$ and $u$ channels
(associated to the three possible ways of connecting up four
participating particles in a Feynman diagram). However, in string
theory the corresponding scattering process has just one contribution,
which embodies within it all three channels. This is a manifestation
of the fact that string scattering is described via a world-sheet,
which can be smoothly deformed at will so that the process resembles
any of the $s,t$ or $u$-channel processes for particle scattering.

Another duality in string theory arises naturally when the spacetime
contains non-contractible circles (a detailed review can be found in
Ref.\cite{Giveon:1994fu}). The string, being an extended
object, can wrap around a circle of radius $R$ giving rise to a
``winding state'' with energy: 
\be 
E~\sim~ n\,\frac{R}{\alpha'} 
\ee
where $n$ is the winding number.  But the spectrum of the theory also
contains familiar states of quantised momentum when the string
propagates as a whole along the compact direction. Such states have a
typical energy:
\be
E~\sim~ \frac{m}{R}
\ee
where $m$ is the quantised mode number.

We see that for a large radius $R$ of the circular direction, the
winding states are heavy due to their spatial extent, while the
momentum states are light. But if $R$ is sufficiently small then the
situation is reversed: winding states are light while momentum states
become heavy due to the uncertainty principle. In fact under the
replacement $R\to \alpha'/R$, the spectrum of winding and momentum
states gets interchanged. Moreover it is known that all interactions of
the string are preserved, with winding and momentum states simply
getting interchanged. Therefore 
\be 
R\to \frac{\alpha'}{R} 
\ee
is an exact symmetry of string theory. This symmetry is called
``target-space duality'' or ``T-duality'' for short.

Both channel duality and T-duality are intrinsic and remarkable
properties unique to string theory.

\subsection{Superstrings and 4d physics}

As a starting point to describe the fundamentals of the real world in
4 space-time dimensions, string theory had all the desirable
properties one might want -- partial uniqueness, gauge and
gravitational interactions, fermions and parity violation. But one had
to derive 4-dimensional physics, specifically the Standard Model at
low energies, from this structure. 

It has long been known in field theory that gauge fields in four
dimensions can be obtained by compactifying a theory of pure gravity
in higher dimensions. This was the Kaluza-Klein mechanism, in which
gauge fields arise as isometries of the compactification manifold.
However, it was shown by Witten\cite{Witten:1983ux} that a
parity-violating spectrum in four dimensions could not be obtained in
this way by starting with a purely gravitational theory in higher
dimensions.  Applied to string theory, this ruled out type IIA and IIB
string theories and focused attention on the type I and heterotic
versions which had non-trivial gauge groups in 10 dimensions. The Type
I open-closed string theory admitted only the $SO(32)$ gauge group and
it gradually became evident that an anomaly-free compactification
starting from this gauge group, under quite general conditions, also
failed to give parity violation in 4d.

That left only the $E_8\times E_8$ gauge group, which was realised
exclusively in heterotic string theory. The low energy field theory of
such a string, compactified to 4 dimensions on a suitable class of
6-dimensional manifolds, was shown in a classic paper of Candelas,
Horowitz, Strominger and Witten\cite{Candelas:1985en} to yield
qualitatively correct phenomenological properties, including parity
violation. The 6-manifolds with favourable four-dimensional
phenomenology were known as ``Calabi-Yau'' spaces. This development
sparked off an explosion of interest in string theory worldwide.

Subsequently a class of simplified compactification models were
discovered\cite{Dixon:1985jw} that make use of ``orbifolds'' rather
than smooth manifolds. Typically orbifolds are singular limits of
smooth manifolds. In appropriate orbifold limits the entire curvature
of a manifold can be concentrated at a discrete set of points while
the rest of the space is flat. This considerably simplifies the
analysis of the spectrum and interactions. Of course such an approach
will not work if the singular curvatures give rise to singular
answers for amplitudes. However it, emerged that under fairly general
conditions string propagation at a singularity is smooth, in part due
to the extended nature of strings.

In hindsight, despite the defining historical role played by the
Candelas et al paper, its central premise -- that a realistic string
theory incorporating and generalising the Standard Model, plus
gravity, was round the corner -- has not proved to be correct.  One of
the fondest hopes implicit in this work, that of starting with an
``almost unique'' theory and obtaining 4d physics via an ``almost
unique'' compactification dictated by elegant mathematical
considerations, is now believed to be far from the truth.  One of the
key issues unresolved issues of the time was how to deal with the
plethora of massless scalars, or ``moduli'' inherent in almost any
compactification. By the time this problem was effectively addressed,
the scenario had changed considerably, as we will see, with the
discovery of many non-perturbative features of string theory.

The study of string compactifications led to the discovery of a new
and unexpected mathematical property\cite{Hori:2003ic} called ``mirror
symmetry''.  This was the occurrence of pairs of 6-manifolds of
Calabi-Yau type with a remarkable property: while the members of the
pair have little geometric resemblance to each other, strings
propagate in precisely the same way on both of them. The interchange
of the two ``mirror'' manifolds is an exact symmetry of string theory.
Thus in the domain of stringy quantum geometry (a concept yet to be
completely unravelled), these manifolds would be completely
indistinguishable from each other. Besides the input to mathematics,
mirror symmetry provides a useful physical result. Upon varying
moduli, examples have been found where a particular manifold undergoes
topology change while its mirror remains smooth. This makes it clear
that topology change can be a perfectly smooth phenomenon in string
theory. It is now understood that mirror symmetry is a sophisticated
version of T-duality, which has made it slightly less mysterious.

\subsection{New insights about gravity II}

Despite the absence of compactification schemes that could reproduce
the real world, some remarkable new properties of space-time emerged
from these investigations. Most of the illuminations stemmed from
T-duality, described above. By relating short (sub-string-scale)
distances to long distances, this challenged the very nature of
geometry. It gradually emerged that a new ``stringy geometry'' is a
more appropriate concept for the way space-time behaves in string
theory. 

An immediate consequence of T-duality was the 
proposal\cite{Brandenberger:1988aj} that when the universe contracts
to a big bang in the far past, one should simply perform a T-duality
when stringy sizes are reached and thereafter the universe will expand
in the new coordinates. More generally, the ability of the string to
effectively resolve singularities and the possibility of topology
change potentially provide important inputs into cosmology.

\section{Non-perturbative string theory}

The latter half of the 1980's saw efforts to systematise (even
classify) two-dimensional conformal field theories, as these were now
understood to correspond to possible string vacua. In particular, it
was hoped that the right compactification 6-manifold, of
``Calabi-Yau'' type, would reproduce the Standard Model in a unified
framework that contained gravity and was UV finite. Apart from many
other complexities, one of the most frustrating problems encountered
in this process was that of massless scalars or moduli fields. To
avoid seriously conflicting with experiment they needed to be
``stabilised'' or given a mass by some mechanism. There were early
hints\cite{Strominger:1986uh} that the inclusion of fluxes over the
compactification space would help, and these turned out correct.
However it proved impossible to stabilise all moduli using known
{\em perturbative} mechanisms.

Therefore it was hoped that non-perturbative effects in string theory
would lift these moduli. Perhaps non-perturbative phenomena could also
rule out some or most of the possible compactifications, whose
ever-growing variety was proving to be an embarrasment.  However, such
effects could not even be studied within the existing formalism of
string theory, essentially a set of rules to produce a perturbative
S-matrix. At this stage string theorists began to seek
non-perturbative information about the theory. A number of routes were
explored and all led to valuable insights.

\subsection{Random matrices and noncritical strings}

The first successful approach to non-perturbative string theory was to
use random matrices to model the string worldsheet as a sort of random
lattice (for a review, see Ref.\cite{Klebanov:1991qa}). In a suitable
continuum limit, the Feynman diagram expansion for the matrix theory
would include all topologies of string worldsheets and one might hope
that results could be extracted beyond perturbation theory. This
approach worked only for vacua of string theory in very low spacetime
dimensions (in particular, one space and one time) and moreover with a
coupling constant that varies along the space dimension\footnote{The
  varying coupling successfully evaded the D=26 consistency condition,
  but unfortunately replaced it with the new condition $D\le 2$.}. The
continuum limit could also be analysed using standard world-sheet
techniques (for a review of the continuum approach, see
Ref.\cite{Mukhi:1991qx}).

We saw that the perturbation series for strings is described 
by a set of surfaces. For each topology, 
we have to sum over all conformally inequivalent
metrics on the surface.
This description is intrinsically 
perturbative in the string coupling.
The matrix approach works instead by discretizing the worldsheet.
\bc
\includegraphics[height=2.5cm]{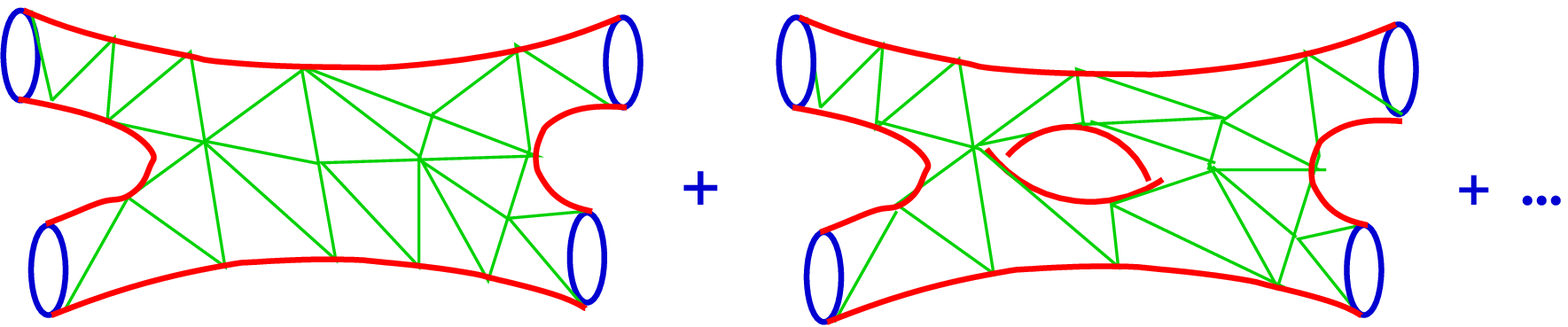}
\ec
Thus the surface is divided into triangles and summing over all ways
to do this implements the sum over all surfaces.
This in turn is done by writing a
random matrix integral, a simple example being:
\be
\int dM\, e^{-\textstyle N\,\tr\,\big(\textstyle\half\, M^2\, + g\,M^3\,\big)}
\ee
where $M_{ij}$ is an $N\times N$ Hermitian matrix.
This can be expanded in a Feynman diagram expansion:
\be
\int dM\, e^{-\textstyle{N\over 2}\,\hbox{tr}\, M^2}\sum_{n=0}^\infty {1\over n!}
\,\left(-gN\, \hbox{tr}\, M^3\right)^n
\ee
which, as usual, is drawn in terms of propagators and
vertices. The propagator for a matrix, $\langle M_{ij}M_{ji}\rangle$,
and the vertex $\tr\, M^3 = M_{ij}\, M_{jk}\, M_{ki}$ are represented
pictorially by:
\bc
\includegraphics[height=1.3cm]{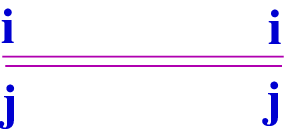}
\qquad\quad${\buildrel{\hbox{and}}\over{\phantom{\strut}}}$\qquad\quad 
\includegraphics[height=2.8cm]{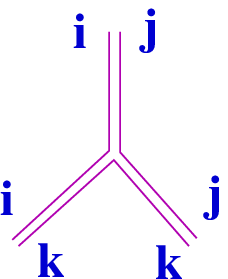}
\ec
respectively.
Connecting propagators and vertices, one constructs graphs that have
cubic intersections, and their dual graphs are then 
triangulated random surfaces.
In the process one finds that the string coupling constant is
effectively: 
\be 
g~\sim~\frac{1}{N} 
\ee where $N$ is the rank of the
matrix.

In these ``non-critical'' strings it proved possible to go beyond just
the finite-order diagrammatic expansion. Here one could compute
scattering amplitudes to all orders in perturbation theory, an
unprecedented achievement, and also identify non-perturbative
effects\footnote{The original non-critical string theories were later
  found to be non-perturbatively inconsistent, but this
  problem was overcome by introducing world-sheet
  supersymmetry\cite{Douglas:2003up}. Therefore in the end this
  development did lead to non-perturbatively consistent and 
well-understood string backgrounds.}. 
The key consequence of these studies was the unexpected discovery that
non-perturbative effects in string theory scale differently from those
in field theory\cite{Shenker:1995xq}. While the latter are typically
of order: 
\be 
\sim~ \exp\left(-\,\frac{1}{g^2}\right)
\label{connon}
\ee
where $g$ is the coupling constant, stringy non-perturbative effects
were found to scale as:
\be
\sim~ \exp\left(-\,\frac{1}{g}\right)
\label{nonconnon}
\ee
which is a considerably larger effect at weak coupling.

This observation was at the root of the discovery, a 
few years later, of D-branes. These objects, viewed as classical
solutions of type II string theory, give rise to effects that
scale as in \eref{nonconnon} because their masses (or, in the
case of instantons, their actions) scale as:
\be
M\sim \frac{1}{g}
\label{nonconmass}
\ee 
This makes them lighter, and their effects correspondingly more
important, compared to the traditional heavy solitons/instantons in
field theory whose masses/actions scale like: 
\be M\sim \frac{1}{g^2}
\label{conmass}
\ee
and which are therefore responsible for conventional non-perturbative
effects as in \eref{connon}.

\subsection{Non-perturbative dualities}

Studies of magnetic monopoles and dyons in quantum field theory reveal
a great deal of similarity between these magnetic objects and the 
usual ``fundamental'' excitations of the quantum field which
carry electric charge. Indeed, the only significant physical
difference between the two classes of objects is that in a weakly
coupled field theory, fundamental electric states are relatively light
(with masses independent of the coupling constant) while solitonic
magnetic states are heavy, their masses scaling as in \eref{conmass}.

Similarities in the dynamics of these objects motivated the proposal
of ``electric-magnetic duality'' -- that there exists a transformation
on a field theory which interchanges the roles of electricity and
magnetism (for a review of this concept and much of the material in this
section, see Ref.\cite{Sen:1998kr}). In order to exchange fundamental
and solitonic objects of the type described above, this duality must
act as: 
\be
\begin{split}
g~&\to~ \frac{1}{g}\\
M(g) ~&\to~ \frac{1}{g^2}\,M\left(\frac{1}{g}\right)
\end{split}
\ee 
Clearly this takes a weakly coupled to a strongly coupled
quantum field theory. So initially it was difficult to imagine how
it could ever be tested.

The resolution came from the ``quantum BPS bound'' or
Witten-Olive\cite{Witten:1978mh} bound. In the original formulation
due to Bogomolny\cite{Bogomolny:1975de} and Prasad and
Sommerfeld\cite{Prasad:1975kr} (BPS) this was a classical lower bound
for the mass of solitons and magnetic monopoles in terms of their
charges and parameters of the theory. When the bound was saturated, it
became easier to obtain classical solutions of the field theory. But
it was realised that generically quantum corrections would affect the
mass, and the bound would therefore no longer be saturated in the
quantum theory. Witten and Olive showed that the situation is very
different in field theories with extended supersymmetry. 

In such theories they noted that the ``central charge'' in the
supersymmetry algebra, evaluated on charged states, evaluates their
electric/magnetic charges.  Moreover when the BPS bound between mass
and charge is saturated, the representations of supersymmetry become
shorter -- the multiplets have a reduced dimensionality. As a typical
example, in a theory with 16 supersymmetry generators a generic state
has a multiplicity of $2^8=256$ but a BPS state has a multiplicity of
just $2^4=16$.  As long as supersymmetry is not violated by quantum
corrections, this ensures -- as an exact operator statement -- that
states saturating the bound classically continue to saturate it at the
quantum level, since quantum corrections cannot continuously deform
a 16-dimensional vector into a 256-dimensional one.

This discovery provided unprecedented control over quantum corrections
in theories with extended supersymmetry. In such cases (the most
famous being ${\cal N}=4$ supersymmetric Yang-Mills theory in 3+1
dimensions) one could now make some definite statements about the
strongly coupled theory. For example the spectrum of BPS states had to be
isomorphic to that at weak coupling, something that is not true
generically without the BPS condition since field theory states can
decay as the coupling changes.

Such considerations led Ashoke Sen in a classic 1994
paper\cite{Sen:1994yi}, to propose a test for strong-weak coupling or
electric-magnetic duality in a compactification of superstring theory
to four dimensions which at low energies (and ignoring $\alpha'$
corrections) reduced to ${\cal N}=4$ super-Yang-Mills theory, He
showed explicitly that the spectrum of this theory fulfils the
requisite conditions for strong-weak duality.  This result sparked off
a wave of interest in non-perturbative duality. Following it, a
variety of dualities were discovered in different field theories,
notably Seiberg-Witten theory\cite{Seiberg:1994rs}, and in compactifications
of string theory\cite{Hull:1994ys,Witten:1995ex}. These dualities were
non-perturbative in nature and could only be argued for using
properties of BPS states as explained above.

The role of duality in string theory was a unifying one. Apparently
different compactifications of different 10-dimensional string
theories were related by conjectured duality transformations for which
stringent tests were proposed.  String theory passed all the tests,
and in each case did so by strikingly different dynamical mechanisms.
During this period, it became more clear than ever that the underlying
structure of string theory was very rigid and constrained, and that
dualities were an intrinsic and deep property built into the theory.

Let us briefly describe how duality operates in uncompactified type II
string theories. In type IIB, the strongly coupled theory is dual to
the weakly coupled one but with an important change: the fundamental
string gets interchanged with a D1-brane (alternatively called
``D-string''). Something quite different happens in the type IIA
superstring -- it gets mapped not onto a weakly coupled string theory,
but into a theory with an extra dimension\cite{Witten:1995ex}. The
large coupling is exchanged for the size of the extra (circularly
compactified) dimension. In the limit of infinite coupling we find
an 11-dimensional theory that has 11-dimensional supergravity as its
low-energy limit. 

Under this duality the string of type IIA string theory reveals itself
to be a membrane of the 11-dimensional theory, which has been dubbed
``M-theory''\cite{Schwarz:1995jq}. This brings 11d supergravity, the
most beautiful and unique of all supergravity theories, into the
framework of strings\footnote{11 is the highest space-time dimension
  in which supergravity exists.}. It also enlarges the scope of string
theory by relating it to a theory where the important excitations are not
strings at all.

Before duality, it was thought that there were five different
superstring theories in 10 dimensions and from them a large variety of
lower-dimensional theories could be obtained via compactification.
However these compactifications now turn out to be
duality transforms of each other, and a path exists (via
compactification, duality and de-compactification) between all the
10-dimensional string theories. Moreover they can all be linked to
M-theory in 11 dimensions.

\subsection{D-branes}

Besides membranes, other extended ``brane''-like solitons are known to
exist in various supergravity theories. These were extensively
explored in the late 1980's and early 90's. The typical strategy for
finding such branes is to postulate that they exist and
saturate the quantum BPS bound. This leads to simpler equations of
motion whose solutions, at least when there is a high degree of
supersymmetry, are guaranteed to solve the full supergravity equations
of motion. In this way, several explicit solutions have been found.
Among such ``brane'' solitons, some of them in type II string theory
carry charges under the Ramond-Ramond gauge fields, a striking fact
given that no states previously known in the theory carried such
charges. Moreover these have tensions that scale according to
\eref{nonconmass} rather than \eref{conmass}, which makes them lighter
than conventional solitons at weak coupling. Finally, they exist in
pairs whose tensions are related by electric-magnetic duality
according to: 
\be 
T(g)= \frac{1}{g^2}\, T\left(\frac{1}{g}\right) 
\ee

Let us examine some of these solitonic brane solutions in a little
detail (see for example Refs.\cite{Polchinski:1998rr,Becker:2007zj}).
For this purpose we will need the supergravity action including the
Ramond-Ramond sector, which was written down in a previous section for
type IIA string theory, retaining only the graviton, dilaton and
1-form Ramond-Ramond gauge field $A_\mu$:
\be S_{\hbox{\litfont
    type}\,\hbox{\litfont IIA}}~=~ {1\over (2\pi)^7 \ell_s^8} \int
d^{10} x\sqrt{-\|G\|}\Bigg[~e^{-2\Phi} \left(R + | d\Phi|^2\right) -
{2\over 8!}| dA|^2\Bigg] 
\ee
Now we start by looking for a classical solution corresponding to a
point particle that is charged under $A_\mu$.
This will be given by a  spherically symmetric gravitational field 
along with an electric flux of  $A_\mu$. Coulomb's law for  field
strengths in 10 dimensions has a $1/r^8$  fall-off, so the field
strength $F_{\mu\nu}=\del_\mu A_\nu-\del_\nu A_\mu$ takes the form:
\be
F_{0r} ~\sim~ {N\over r^8},~r\to\infty
\ee
where we anticipate that there will be  $N$  quantised units of this
flux.

The gravitational field is specified by writing the metric:
\be
ds^2 = -\left(1+{r_0^7\over r^7}\right)^{-{1\over 2}}\, dt^2 ~+~ 
\left(1+{r_0^7\over r^7}\right)^{1\over 2}
\, \sum_{a=1}^9 dx^a dx^a
\label{dzmet}
\ee
where  $r=\sqrt{x^a x^a}$.
This is like an  extremal Reissner-Nordstrom black hole  in
10 dimensions (in these coordinates the horizon is at  $r=0$).
To complete the solution we have to specify the  dilaton  
and  gauge potential:
\be
\begin{split}
e^{-2\Phi} &= e^{-2\Phi_0}
\left(1+{r_0^7\over r^7}\right)^{-{3\over 2}}\\
A_0 &= -\half \left[\left(1+{r_0^7\over r^7}
\right)^{-1}-1\right] 
\end{split}
\label{dzdilpot}
\ee
where we recall that  $g_s = e^{\Phi_0}$.

We can compute the mass of this object from the classical solution,
and the result comes out to be: 
\be M = {1\over
  d\,g_s^2\,\ell_s^8}\,\left(r_0\right)^7,\qquad 
\ee 
where $d$ is a constant. Using the Dirac quantisation condition one
can argue that this object can only occur in integer multiples of
a minimally charged object, where the integer is:
\be
N = {1\over d\,g_s\,\ell_s^7}\,(r_0)^7
\ee
This can be thought of as the charge of the object in units where the
minimum charge is unity. The supergravity solution is valid only
when $N$ is large, i.e. $r_0\gg \ell_s$, otherwise the curvatures will
be large and we are not entitled to use the lowest-order action in
$\alpha'$.

From the above formulae we see that: 
\be 
M = {1\over g_s \ell_s} \, N
\ee 
Now, just using supersymmetry one can prove that states charged
under $A_\mu$ in this theory obey a mass bound: 
\be 
M \ge {1\over g_s
  \ell_s} \, N 
\ee
Since the above soliton saturates this bound, it must correspond to a
stable particle state in the theory. Also we see clearly that although
there are no particles in the perturbative string spectrum carrying RR
charges, the soliton exhibited above precisely carries such charges.

So far the physical interpretation has been quite conventional. This
changed when Polchinski, in a landmark paper\cite{Polchinski:1995mt}
in 1995, observed that such branes admit an alternate description as
the end points of open strings. The idea is that within a closed
string theory, there are dynamical objects with the property that open
strings can end on them. In fact the open strings ending on this
object provide an alternative description of the object itself. 

To end on a fixed object, open strings must have Dirichlet boundary
conditions in all directions transverse to the object.  These
partially break the Lorentz invariance and supersymmetry of the
underlying closed string background. This is understandable because
the brane in question is an excited state of the closed string theory
and therefore (like any excited state) should spontaneously break some
of the original symmetries.

As a special example corresponding to a point-like particle, consider
an open superstring with Dirichlet boundary conditions on all 9 space
directions on each of its two endpoints. To be specific, we restrict
both ends to lie at the origin in 9-dimensional space.  Being thus
nailed down, such a configuration clearly has no centre-of-mass degree
of freedom.  Therefore the excitations of this open string are all
bound to the location of the string end-points.  In this situation the
effective field theory for the open string is not a 10d field theory
at all, but just quantum mechanics on a ``world-line'' fixed at the
origin of space. The claim is that this quantum mechanics provides an
alternate description of the pointlike soliton described above in
Eqs.(\ref{dzmet}),(\ref{dzdilpot}). 

Let us see why 9 Dirichlet boundary conditions describe a particle.
These boundary conditions clearly break Lorentz as well as translation
invariance in 10d. However, $SO(9)$ rotational invariance around the
origin is preserved: \be SO(9,1)\to SO(9) \ee Moreover, in the
world-line theory the would-be gauge field $A_\mu$ (which would have
been present had the ends of the string been free to move) is
re-interpreted as a (non-dynamical) gauge field $A_0$ along with nine
scalar fields $\phi^a$. Now in field theory, a particle state breaks
translational invariance, since translations move the particle to
another point. But it preserves rotational invariance around the
location of the particle. We see that the endpoint of the Dirichlet
open string described above has the right properties to be a dynamical
particle. This interpretation provides a nice interpretation for the 9
scalar fields on its worldline: they would be the 9 spatial
coordinates of this particle!

With this interpretation, the mass and charge of the string endpoint
(which we now refer to as a ``D-particle'') can be computed within
string theory\cite{Polchinski:1995mt}. The result is that the
D-particle carries precisely one unit of charge under the
Ramond-Ramond gauge field $A_\mu$.  Moreover its mass is:
\be 
M = {1\over g_s\ell_s} 
\ee 
These properties are consistent with, and support, the notion that the
open string endpoint describes a unit-charged version of the RR
soliton exhibited as a classical solution in
Eqs.(\ref{dzmet}),(\ref{dzdilpot}).

The above discussion can be generalised to the case where the string
has Neumann boundary conditions at each end in some of the 9 space
directions, say $1,2,\cdots,p$, and Dirichlet conditions in the
remaining $9-p$ directions.  \bc
\includegraphics[height=3.5cm]{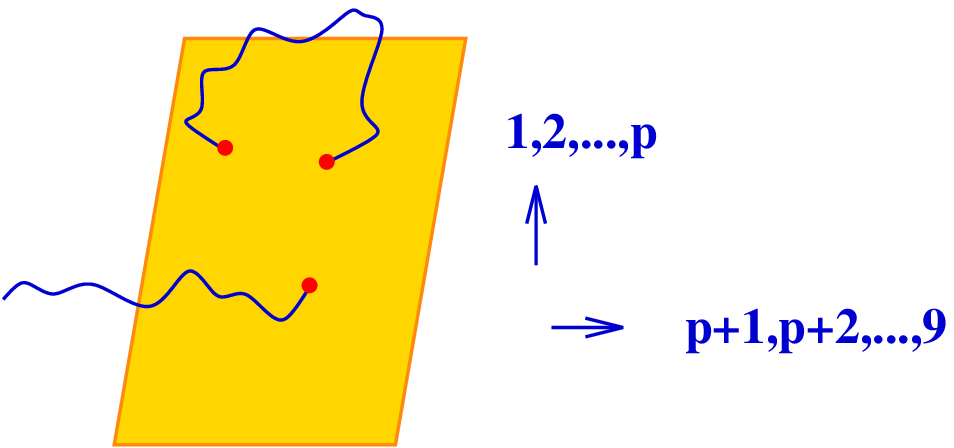}
\ec This defines a $p$-dimensional hypersurface in spacetime.
Generalising the ``D-particle'', such a wall is called a
``D$p$-brane''. We find that the massless states are now a photon
$A_\mu$ in $p+1$ dimensions, as well as $9-p$ scalar fields $\phi_a$
(plus, of course, fermions). Clearly the low-energy effective field
theory on a D$p$-brane is a $(p+1)$-dimensional field theory.  \bc
\includegraphics[height=2.5cm]{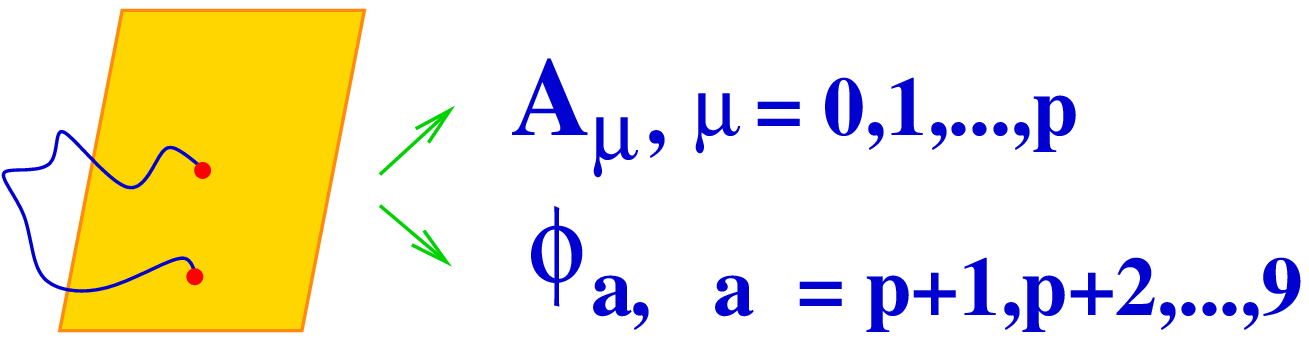}
\ec
In particular, there is   one scalar field for each direction
transverse to the brane.
As before, the vacuum expectation value of these scalars are naturally
interpreted as the transverse locations of the branes.

It was perhaps misleading to refer to a D-brane as a ``wall'' as we
did above. Since it has a fixed tension (which matches that of its
dual description as a soliton) it can be deformed in all possible ways
simply by providing appropriate amounts of energy. So it is simply an
extended dynamical object. A planar D-brane is of course the most
symmetric allowed configuration and therefore also the simplest to
find as a classical solution.

There are stable D-branes charged under each of the RR fields of type
IIA/B supergravity. Over the years, each type of brane has provided
rich new insights into quantum field theory in the corresponding
space-time dimension. By far the most profound insight comes from the
AdS/CFT correspondence which we discuss in the following section. Here
we briefly describe a simpler physical insight that is a precursor to
the correspondence: the origin of non-Abelian gauge symmetry.

For this, let us assemble a collection of N parallel planar D-branes.
As gravitating objects one might expect them to attract each other,
but due to supersymmetry there are extra exchange forces between them
besides the gravitational one, and these neatly cancel so this
configuration is stable.
\bc
\includegraphics[height=3cm]{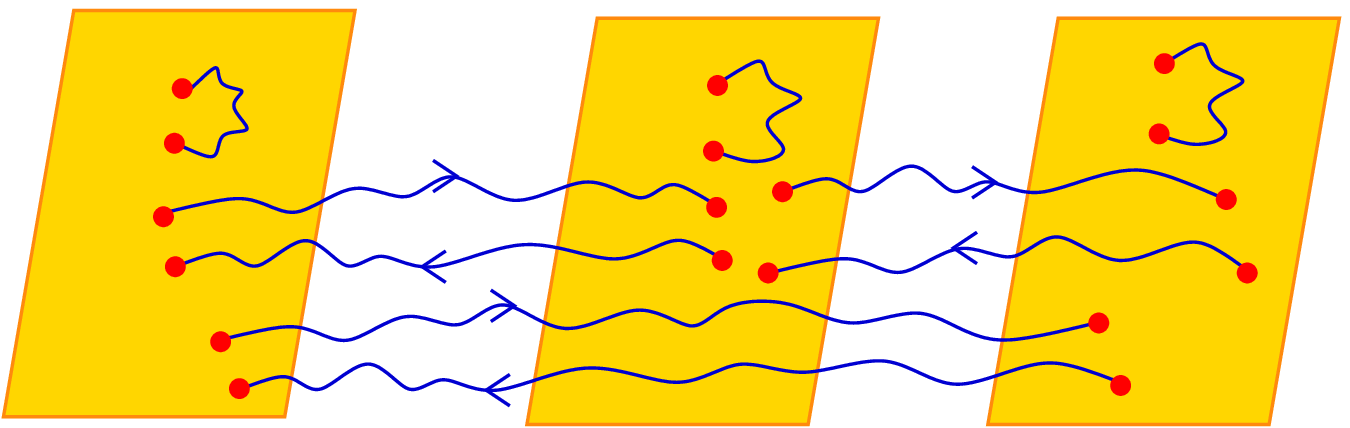}
\ec 
In this array, an open string can start on any one of the branes
and end on any other. Thus there are $N^2$ species of open strings.
That means their lowest excitations, each one of them a massless gauge
field, can be collected into an $N\times N$ matrix $A_\mu^{\alpha\beta}$.

Let's consider the simplest example, a pair of D3-branes of type IIB
string theory:
\bc
\includegraphics[height=3cm]{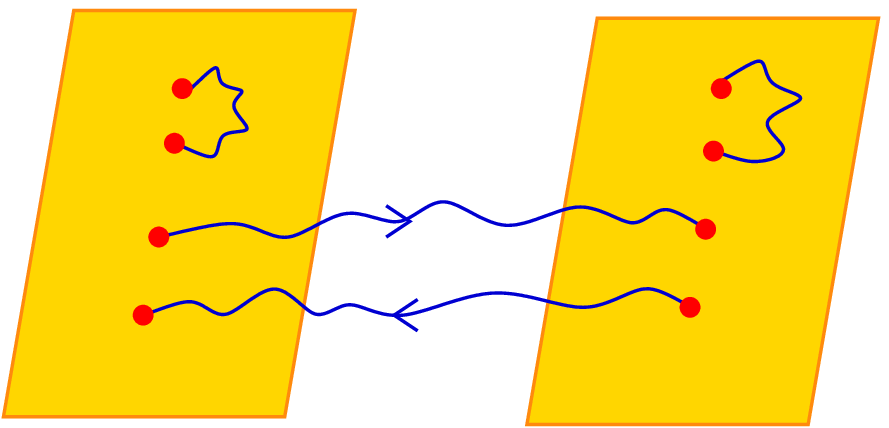}
\ec 
We see that there are four species of strings. Of these, two are
localised on individual branes, so they clearly represent the Abelian
gauge field for that brane. Together, these two strings provide
$U(1)\times U(1)$ gauge fields. However the two strings stretching
across the branes are electrically charged under $U(1)\times U(1)$.
Thereby they provide the extra gauge fields to enhance: \be U(1)\times
U(1)\to U(2) \ee If the two D-branes are precisely coincident, then
the strings stretching from one to the other can shrink to zero length
under their own tension. At this point, all the four gauge fields are
massless. If we now separate the branes, two of the four strings
acquire a minimum length and therefore a classical energy. So the
corresponding gauge fields must be massive.  Since transverse motion
of the branes is represented by giving a VEV to the transverse scalar
fields, this is string theory's geometric realisation of the Higgs
mechanism!

Quantising the $N^2$ strings on a stack of $N$ D3-branes and
performing amplitude calculations, one finds that
the low-energy effective theory is
the   Yang-Mills theory  of a  U(N) \ gauge field
 $A_\mu^{\alpha\beta},~\alpha,\beta=1,2,\ldots N$, coupled
to scalars $\phi^a$ and fermions $\psi^A$ in the adjoint 
representation of  U(N), with the action:
\be
\begin{split}
\cL &= \tr\,\Big\{-{1\over 4 g_{YM}^2}F_{\mu\nu}F^{\mu\nu}
-\half D_\mu \phi^a D^\mu \phi^a 
-{g_{YM}^2\over 4}[\phi^a,\phi^b]^2 \\
&\qquad\qquad\qquad + {i\over 2}{\bar\psi}^A \gamma^\mu D_\mu
\psi^A - g_{YM} {\bar\psi}^A\Gamma^a_{AB} [\phi^a,\psi^B] ~\Big\} 
\end{split}
\label{neqfour}
\ee
where $a,b=1,2,\cdots,6;~ A,B=1,2,\cdots,4$ and $g_{YM} = \sqrt{g_s}$.
All fields in the action are matrices.  This action has the maximal
supersymmetry allowed for a gauge theory in 4 dimensions, namely
${\cal N} =4$ supersymmetry, and is completely dictated by
supersymmetry.

A special property of this theory, arising from the constancy of the
dilaton in the classical solution, is that it is conformally
invariant. Its $\beta$-function in fact vanishes to all orders in
$g_{YM}$, which renders it scale invariant. As often happens in field
theory, scale invariance automatically gets promoted to conformal 
invariance. One consequence is that supersymmetry is enhanced: by
commuting special conformal transformations with the usual 16
supersymmetries, one generates 16 new supersymmetries.  Among D$p$-brane
theories for $p=1,2,\cdots 9$, this is the only conformally invariant
theory.

 Amplitude calculations with multiple D-branes explicitly reveal the
 famous ``three-gluon''\ and ``four-gluon''\ interactions, the
 signature of Yang-Mills theory: 
\be 
\tr\, \del_\mu A_\nu
 [A^\mu,A^\nu],\qquad \tr\,[A_\mu,A_\nu][A^\mu,A^\nu] 
\ee 
In addition, they correct the Yang-Mills action written above with
$\alpha'$ corrections involving higher derivatives of the fields.

Besides introducing non-Abelian gauge symmetries into string theory,
D-branes also help reformulate familiar notions from field theory and
mathematics in a new way. This leads to several novel insights about
gauge theory and gravity. Indeed the very conception of string theory
and its role as a theory of quantum gravity has undergone a fairly
radical change after Polchinski's discovery. Being intrinsically
solitonic, D-branes are non-perturbative objects like magnetic
monopoles in field theory (and heavy at weak coupling like these
monopoles). Nevertheless, perturbative techniques in open string
theory can be brought to bear on their dynamics.  In this way one
acquires the power to reliably analyse non-perturbative excitations.
String theory is no longer just a theory of strings, but of extended
branes of all kinds, and their gravitational dynamics is, via open
strings, inextricably linked to gauge dynamics.

\subsection{M-theory}

As already mentioned above, dualities led to the discovery of a new
theory called M-theory.  It has similar properties to string theory in
that it possesses extended objects. However it has no stable strings.
It is well-defined only in 11 dimensions, though like string theory it
can be reduced to any dimension $d<11$ by compactification. One reason
to believe M-theory is fundamental is that, as noted above, 11 is the
highest allowed number of dimensions for a consistent supersymmetric
theory.

A significant difference from string theory is that M-theory has no
dilaton and consequently no perturbative expansion. However, on
compactifying the low-energy action of M-theory from 11d to 10d, we
recover the low-energy action of type IIA string theory.  Indeed, the
dilaton of string theory emerges as a scalar mode of the
11-dimensional metric.  This is one more reason why M-theory seems
more basic than string theory.

M-theory is not very well-understood, but even at the level at which
we understand it, it ``explains'' many interesting features of string
theory. In string theory there are several $p$-form fields and the
objects charged under them are strings and branes. Just considering
stable objects, type IIA string theory for example has three different
$p$-form fields and as many as seven different types of strings and
branes. However M-theory has a single 3-form field and a basic brane,
a membrane (extended in two space dimensions) electrically charged
under it. There is also its dual, a 5-brane that is magnetically
charged under the 3-form field (for details, see for example
Ref.\cite{Becker:2007zj}).

On compactifying M-theory on a circle, these two branes can wrap -- or
not wrap -- the circle and as a result, they give rise to all the
branes of string theory\cite{Aharony:1996en,Aharony:1996xr}, including
the fundamental string. 

Since it is the membrane of M-theory that gives rise to the
fundamental string upon compactification, we may guess that this is
the closest to a fundamental object in M-theory\footnote{In fact, M
  stands for Membrane among other things.}. Given that there are no
strings in the theory, a fascinating question is what generates
interactions between parallel membranes.  The answer is that when one
has parallel membranes then they can be smoothly be connected by other
membranes into a single larger object. After compactifying the theory
on a suitable circle, this object reduces to a configuration of type
IIA D2-branes and strings running between them. These considerations
have led to a new understanding, though still incomplete, of membrane
interactions in
M-theory\cite{Bagger:2007jr,Mukhi:2008ux,Aharony:2008ug}.

\subsection{Black Holes}

Being a consistent theory of quantum gravity, string theory can be used
as a testing ground to delve into the nature of black holes and the
various potential paradoxes surrounding these objects.  The nature of
black hole evaporation, the entropy of black holes and the possibility
of information being lost in black holes were all issues that had been
widely discussed for many years. With the discovery of string
dualities and D-branes, reliable and precise tests became possible.

The basic idea is to consider string states carrying a fixed set of
charges at weak coupling, where they can be thought of as
string/D-brane excitations. The number $\Omega$ of microscopic states
of these excitations can be counted and their logarithm is the
``statistical entropy'' of the system. By contrast, at strong coupling
these states fall inside their own Schwarzschild radius and turn into
a black hole\cite{Susskind:1993ws,Horowitz:1996nw}.  The arguments of
Beckenstein and Hawking can then be used to compute the entropy of
these black holes in terms of their famous area formula. Thus one had
two independently computed quantities:
\be
S_{\hbox{statistical}}=\ln \Omega,\qquad 
S_{\hbox{black hole}}= \frac{A}{4}
\ee
(both evaluated in suitable units where all dimensional constants are
set equal to 1).  In principle there might have been little connection
between these quantities, arising respectively from the spectrum of
states at extremely weak and strong coupling. However in
supersymmetric theories it is possible to relate the two extremes
under some circumstances due to the remarkable quantum BPS property
alluded to earlier.

In a major breakthrough, Strominger and Vafa\cite{Strominger:1996sh}
in 1996 computed both quantities for dyonic extremal charged black
holes in 5 dimensions\footnote{The dimensionality was merely a
  convenience, and  analogous results were soon obtained for extremal
  black holes in four dimensions.}. The black-hole calculation is
standard, requiring knowledge of the black-hole metric which can
then be integrated over the horizon to obtain the area. However the
microscopic calculation, as well as the reason why it should be relevant,
is special to superstring theory. This computation is based on
a configuration of intersecting D-branes of different dimensionalities
wrapped around different cycles of the space-time manifold. The system
carries fixed electric and magnetic black hole charges $Q_e,Q_m$, and
is described by a field theory on the uncompactified part of the
D-brane world-volume, which is two dimensional and also turns out to be
conformally invariant. Standard techniques based on the
Virasoro algebra then allow one to estimate the degeneracy of states
of this theory, at least in the limit that some of the charges are
taken large (recall that the black hole picture also makes sense for
large charges).

If $Q_e, Q_m$ are the (integer) electric and magnetic charges of the
system, the result of Ref.\cite{Strominger:1996sh} is:
\be
\bsp
S_{\hbox{statistical}}&=2\pi\sqrt{Q_m\left(\half Q_e^2+1\right)},
~~~Q_e~\hbox{fixed},~Q_m~\hbox{large}\\[3mm]
S_{\hbox{black hole}}&= 2\pi\sqrt{\half Q_m Q_e^2}
\qquad \qquad\quad Q_e,Q_m~\hbox{both large}
\end{split}
\ee
Note that the second formula is valid in a more limited range of
parameters than the first, so agreement is expected only upon taking
$Q_e$ large in the first line. In this limit the two
expressions agree perfectly, including the numerical coefficients.
This calculation strongly supports the notion that a black hole is
fundamentally a statistical ensemble and its entropy is due to the
microstates that make it up. 

It also suggests a class of generalisations: if one could compute both
quantities in the above equation for finite charges, or at least up to
some finite order in an expansion in inverse charges, one should find
agreement in each order.  On the statistical side this requires
greater control over the conformal field theory, allowing the
computation of the degeneracy of states to higher accuracy as an
expansion in large charges. On the black hole side the relevant
quantity is the Wald entropy, defined via a formula that generalises
the Beckenstein-Hawking formula to theories with higher-derivative
terms in the action. It is the Wald entropy rather than the
Beckenstein-Hawking entropy that satisfies the second law of
thermodynamics in general theories. In string theory we have seen that
the higher-derivative terms are uniquely fixed by the background and
therefore the Wald entropy sensitively probes the stringy origin of
the gravity action. Phenomenal agreement has been found between the
corrections to $S_{\hbox{statistical}}$ and to $S_{\hbox{black hole}}$
and the field of ``precision counting'' of black hole states has now
come into being (for a review see Ref.\cite{Sen:2007qy}, while more
recent results can be found for example in
Ref.\cite{Dabholkar:2010rm}).

\subsection{New insights into gravity III}

The triangulation of Riemann surfaces by random matrices correctly
reproduces, in the continuum limit, what we expect from summing over
continuous surfaces, the latter being a calculation in
two-dimensional gravity. It is found explicitly that in the domain of
overlap, the continuum approach agrees with the results from matrices.
However the matrix approach is actually much more powerful, allowing
the computation of amplitudes to all orders in the string
coupling. These insights about two-dimensional gravity arose as a kind
of by-product of string theory.

But it is D-branes that have brought about the most major revolution
in our understanding of quantum gravity through string theory.  They
are massive solitonic objects, excitations of the closed-string
theory, so they of course gravitate and the metric of space-time
around them is precisely known via supersymmetry and BPS equations. On
the other hand they are described by open strings on their
world-volume and the low-energy limit of this theory is a gauge theory
of Yang-Mills type. These two descriptions of D-branes amount to a
duality, crucial to the black hole entropy calculations sketched
above: the gravitating description of D-branes provides the black-hole
entropy while the open-string description of the same objects counts
the microscopic degrees of freedom. This is an example of what one
could call ``open-closed duality''. The most profound example of such
a duality is the AdS/CFT correspondence, to be discussed in the following
section.

\section{AdS/CFT correspondence}

\subsection{Precise statement via heuristic derivation}

The discovery of D-branes made it natural to consider a system of open
string states inside a closed-string theory. A particular example of
such a system is a stack of $N$ D3-branes, on whose volume, as we have
seen, the open strings generate an ${\cal N}=4$ supersymmetric gauge
theory in four space-time dimensions. 

In 1996-7, this system was extensively studied in the limit of large
$N$. Following some earlier key
results\cite{Klebanov:1996un,Klebanov:1997kc},
Maldacena\cite{Maldacena:1997re} observed that on the one hand, at low
energies this is a conventional gauge theory (even though
supersymmetric and with a gauge group of large rank). On the other
hand a brane, viewed as a solitonic object, deforms the spacetime
around it so that the geometry near the brane is very different from
the flat spacetime far away. For $N$ D3-branes, this local geometry
has the form of five-dimensional Anti-deSitter spacetime AdS${}_5
\times$ a 5-dimensional sphere. In a path-breaking work using
symmetries and dynamical arguments, Maldacena proposed that the two
descriptions are exactly equivalent, so that a four-dimensional gauge
theory is the same as a closed string theory (including gravitation!)
in a particular 10-dimensional spacetime.  This, the latest and
perhaps most dazzling of string dualities, has revolutionised our
notions about gravity and gauge theory: previously thought of as two
totally distinct types of theories, we must now accept that under
certain circumstances they can be one and the same theory. Moreover
the correspondence is ``holographic'' in the sense that gravity
degrees of freedom are encoded in a gauge theory that lives in a lower
number of dimensions.

In concrete terms, consider the Lagrangian of type IIB supergravity
(that arises as the low-energy limit of the type IIB string). This
time we include the self-dual 4-form Ramond-Ramond gauge field that is
present in this theory: 
\be 
S_{\hbox{\litfont type}\,\hbox{\litfont
    IIA}}~=~ {1\over (2\pi)^7 \ell_s^8} \int d^{10}
x\sqrt{-\|G\|}\Bigg[~e^{-2\Phi} \left(R + | d\Phi|^2\right) - {2\over
  5!}| dD^+|^2\Bigg] 
\ee
(technically the self-duality condition makes
$| dD^+|^2$ vanish, so we impose that condition after computing the
equations of motion). Now we can write the metric for a classical
solution corresponding to a ``black
3-brane'': 
\be ds^2 = \left(1+{R^4\over
    r^4}\right)^{-{1\over 2}}\, \Bigg(-dt^2 + \sum_{i=1}^3 dx^i
dx^i\Bigg) + \left(1+{R^4\over r^4}\right)^{1\over 2} \,
\sum_{a=1}^6 dx^a dx^a 
\ee
where $r=\sqrt{x^a x^a}$. Here $R$ is a constant parameter of the
solution (not to be confused with the scalar curvature!). 

The dilaton is constant in this solution: \be e^{-2\Phi} =
e^{-2\Phi_0} \ee while the 4-form potential is: 
\be D^+_{0123} = -\half \left[\left(1+{R^4\over
      r^4}\right)^{-1}-1\right] 
\ee
The quantised charge $N$ of this solution is easily found to be:
\be
N = \frac{R^4}{4\pi g_s\ell_s^4}
\ee
Notice  that in 10 dimensions, a 3-brane is enclosed by a 
5-sphere  and the integral of the field strength  $dD^+$ 
over  this 5-sphere measures the total charge  $N$.

We now examine physics of a  test particle  in this field.
The coefficient of $-dt^2$\ tells us there is a  redshift 
between the energy measured at some radial distance $r$ and at $\infty$:
\be
E_\infty = \left(1+{R^4\over r^4}\right)^{-{1\over 4}}E_r
\ee
This means that a given object near $ r\to 0$\ has a very small
energy when  measured from infinity.
Define: 
\be U \equiv {r\over \ell_s^2}
\ee
which is a spatial
coordinate with dimensions of energy. 
Then, multiplying through by $\ell_s$, we find:
\be
E_\infty \ell_s =  \left(1+{4\pi g_s N\over 
(U\ell_s)^4}\right)^{-{1\over 4}}E_r\ell_s
\ee
This shows that from the point of view of an observer at
infinity, low energy $ E_\infty \ell_s\ll 1$\ means:
\be
U\ell_s\ll 1~~\hbox{or}~~ E_r\ell_s\ll 1
\ee
Indeed this low energy limit can be thought of as $\ell_s\to 0$
with  energies held fixed.

In the first regime, the metric of the D3-brane becomes:
\be
ds^2 = \sqrt{4\pi g_s N}\,\ell_s^2
\Bigg[ {U^2\over 4\pi g_s N}(-dt^2 + dx^i dx^i)
+ {dU^2\over U^2} + d\omega_5^2
\Bigg]
\ee
This is the metric of the spacetime AdS${}_5\times S^5$. There is also
an RR field strength in the classical solution. 
The second regime instead describes states of small proper energy in
units of $\ell_s^{-1}$. Such states correspond to the $\ell_s\to 0$\
limit of supergravity, which is free in this
limit.

Next one uses the dual description of D3-branes as open-string endpoints.
In this description, the system is described by an effective action
for  open strings  plus an action for  closed strings  plus an action
describing  open-closed couplings:
\be
S=S_{\hbox{open}} + S_{\hbox{closed}} + S_{\hbox{open-closed}}
\ee
Taking $ \ell_s\to 0$\ keeping energies fixed, the closed-string
part (supergravity) becomes free and the
open-closed couplings also vanish.
Finally, in the open-string part, the higher-derivative terms
disappear since they are proportional to powers of $ \ell_s$.
The surviving action is the $ {\cal N}=4$\ supersymmetric Yang-Mills
field theory written in \eref{neqfour}, with gauge group $ U(N)$\ and
coupling constant $ g_{YM}=\sqrt{g_s}$. 

Thus comparing the two sides we see that each one has a free
supergravity action, which can be  equated.
The remaining part, which can also be equated, is
(i) string theory in the curved background AdS${}_5\times
S^5$, and (ii) $ {\cal N}=4$\  supersymmetric
Yang-Mills field theory.
The  AdS/CFT correspondence  is the conjecture that  these
two theories are the same.

Unlikely as it may seem, this conjecture says that string theory in a
particular bulk spacetime is equal to a conformal-invariant field
theory in a (conformally) flat space-time that corresponds to
the boundary of the original bulk space-time.  Moreover the
dimensions of the two theories are 10 and 4 respectively.

The above was a heuristic derivation following
Ref.\cite{Maldacena:1997re}, but the AdS/CFT correspondence has not
yet been rigorously proven. It has been tested in many different ways
though, and some of these tests are briefly described below.

\subsection{Matching symmetries: isometries}

To test the AdS/CFT correspondence, one can first check that the
symmetries match on both sides.  The isometries of AdS${}_5\times S^5
$ are: 
\be
\begin{split}
\hbox{AdS}_5:\quad & SO(4,2)\\
S^5: \quad & SO(6)
\end{split}
\ee

On the other side of the correspondence we have a conformally
invariant field theory, $ {\cal N}=4$\ supersymmetric Yang-Mills
theory. It clearly possesses $ SO(3,1)$\ symmetry, namely Lorentz
invariance. Another symmetry we see right away is global $ SO(6)$\
invariance which rotates the six scalar fields $ \phi^a$\ that
describe transverse motions of the D3-brane.  The remaining desired
symmetries arise from the fact that whenever a field theory has
conformal invariance, this symmetry combined with Lorentz invariance
gives rise to an enhanced symmetry group: \be SO(d,1) \to SO(d+1,2)
\ee Thus indeed, $ {\cal N}=4$\ supersymmetric Yang-Mills theory has $
SO(4,2)\times SO(6)$\ symmetry, just like the isometries of 
AdS${}_5\times S^5$.  This is a successful test of the AdS/CFT
correspondence. 

Another test follows from matching supersymmetry. Closed superstrings
propagating in flat spacetime have $ {\cal N}=2$\ supersymmetry in 10
dimensions.  The supercharges have $ 16$\ components each, making a
total of $ 32$\ components. The only other 10 dimensional space-time
with the same number of supersymmetry charges is AdS${}_5\times S^5$. $
{\cal N}=4$\ SYM theory has $4$ spinor supercharges, each with $4$
components.  Therefore there are apparently just $ 16$\
supersymmetries. However, as we mentioned earlier, taking the
commutator of special conformal transformations with supersymmetries
gives rise to a new set of supersymmetries, also $ 16 $\ in number.

Thus at the end, both sides  have $ 32 $\ supersymmetries. In fact
one can show that:
\be
SO(4,2)\times SO(6)\times \hbox{susy}~\subset~
SU(2,2| 4)
\ee
where the RHS is a particular  super-algebra, which is a symmetry
of both sides of the AdS/CFT correspondence.

\subsection{Parameters and gravity limit}
\smallskip

The proposed duality is so nontrivial that, beyond symmetries, it is
not immediately obvious how to test it or use it.
One major obstacle is that string theory on AdS${}_5\times S^5$\
has Ramond-Ramond flux. We do not know how to study strings propagating
in the presence of such backgrounds.
Thus we are forced to restrict ourselves to the  low-energy
effective action  of string theory, namely  supergravity.
This is valid in the  weakly curved  case:
\be
R \gg \ell_s
\ee
which amounts to:
\be
\lambda\equiv g_{YM}^2 N\gg 1
\ee

At the same time we must restrict to  tree-level, since 
supergravity is a non-renormalisable theory so loop diagrams will not
make sense. Therefore we must have:
\be
g_s \ll 1~\Longrightarrow~ g_{YM}\ll 1
\ee
It follows that the gauge theory must have $ N\gg 1$. Indeed, the
behaviour of gauge theories at large $ N$ (and  the simple example of
random matrices discussed above) were among the earliest
indications that field theory is related to string theory!
\smallskip

\subsection{Gravity-CFT dictionary}

There is a precise dictionary between gravity variables and gauge theory
variables, that is known explicitly in many cases. The general
proposal\cite{Witten:1998qj,Gubser:1998bc} is that to each
gauge-invariant operator ${\cal O}(x^\mu)$\ in the SYM theory, there
corresponds a field $\phi(x^\mu,U)$\ in supergravity such that:
\be 
\bsp
\Big\langle
\hbox{exp}\left(\int d^4 x\, J(x^\mu)\,{\cal
    O}(x^\mu)\right)\Big\rangle_{\hbox{\small gauge theory}}&= {\cal
  Z}_{\hbox{\small supergravity}}
\Big(\phi(x^\mu,U\Big)\bigg|_{\phi(x^\mu,U\to\infty)=J(x^\mu)}\Big)\\
&
\end{split}
\ee 
Here the LHS is a gauge theory correlation function in 4 dimensions.
The RHS is the gravity partition function evaluated on 5-dimensional
fields $\phi(x^\mu,U)$\ in AdS${}_5$, but with their values constrained to be
equal to the source $J(x)$\ on the boundary of AdS${}_5$.
We can generalise this to supergravity fields that depend on the 
$S^5$\ coordinates, by Fourier decomposing them on $S^5$\ and treating
each Fourier mode as an independent field on AdS${}_5$.

As a relatively simple example, consider the marginal operator which
changes the gauge theory coupling constant. This is just the entire
Lagrangian of the gauge theory!  In the gravity dual the corresponding
field in 5 dimensions is the dilaton operator $\Phi(x^\mu,U)$. Its
value on the boundary of AdS${}_5$\ determines the coupling of the gauge
theory.  Thus in this case the correspondence is: 
\be 
\bsp
\hbox{Operator in gauge theory} \qquad
\Longleftrightarrow&\qquad\hbox{Field in supergravity}\\
-{1\over 4} \tr\, F_{\mu\nu}F^{\mu\nu}(x^\mu) + \cdots
\qquad\Longleftrightarrow & \qquad \Phi(x^\mu,U)
\end{split}
\ee

We see that the extra holographic dimension on the gravity side is the
radial direction $U$. This can be shown to correspond to an energy
scale in the field theory. Conformal invariance of the field theory
is natural in this interpretation. The dilaton background is constant
in the AdS classical solution, therefore in particular it is
independent of $U$. Therefore the dual field theory is independent of
energy scale, which in turn implies conformal invariance.

If we want to generalise AdS/CFT to have a scale-dependent theory like
QCD on the gauge theory side, then the dual spacetime must be
different from AdS in the interior, and the dilaton must be a
non-trivial function of $U$.

\subsection{An application of the correspondence}

Since one side of the correspondence is classical gravity, which is
relatively easy to study, we can use it to deduce properties of
quantum gauge theories at large $N$\cite{Witten:1998zw}.  In nature we
do not want to know about the gauge theory called ${\cal N}=4$
supersymmetric Yang-Mills, but about Quantum Chromodynamics, however
at finite temperature the ${\cal N}=4$ SYM can be shown to resemble
Quantum Chromodynamics in some ways.

We start by placing the gauge theory not on $R^{3,1}$\ but on $
S^3\times S^1$.  This in particular requires us to make the theory
Euclidean, corresponding to finite temperature. If $ \beta$\ is the
radius of $ S^1$, then the temperature is: \be T = {1\over \beta} \ee
We also define the radius of $ S^3$\ to be $\beta'$.

Conformal invariance then tells us the theory depends only on the
dimensionless ratio $ \beta/\beta'$.  It has been
shown\cite{Witten:1998zw} that there are two candidate gravity duals
to this theory. One is a spacetime called thermal AdS (like AdS${}_5$\
but at finite temperature). The other is a Schwarzschild black hole
which asymptotically becomes AdS.  Which of these two is the correct
gravity dual depends on the temperature, more precisely on
$\beta'/\beta$. At small values of this parameter (low temperature)
the thermal AdS dominates the path integral. At high temperatures
instead it is the AdS black hole.

Now the gravity description can be used to compute the entropy in each
case. At low temperatures it is found that: 
\be 
S~\sim~ 1 
\ee
while at high temperatures, the Bekenstein-Hawking formula for black
holes gives us: \be S~\sim~R^3\times R^5~\sim~R^8\sim~N^2 \ee The jump
from one to another AdS dual of the field theory as we vary
temperature is a phase transition, and is interpreted as the
deconfinement phase transition of the gauge theory! The numbers fit
beautifully with the fact that at low temperature the gauge theory has
only singlet states but at high temperature the gluon degrees of
freedom, which are $N^2$ in number, are liberated\footnote{See
  Ref.\cite{Gubser:1996de} for early results in this direction. The 
present discussion may
  seem somewhat confusing given that ${\cal N}=4$ supersymmetric
  Yang-Mills theory does not exhibit confinement. This and other
  subtleties are explained in
  Ref.\cite{Witten:1998zw}.}.

We see the power of the AdS/CFT correspondence in extracting  analytic
information about confinement. Though the gauge theory in this
discussion is not a realistic one, there have been many attempts to
generalise the above discussion to more realistic confining examples
and this remains an active direction of research.

\subsection{Gravity and fluid dynamics}

Gravitational theories which allow asymptotically AdS space-times can
be truncated to pure Einstein gravity with a (negative) cosmological
constant. This implies a certain universality property for the dual
gauge theory on the boundary, which in turn can be encoded in the
dynamics of the gauge-invariant fields, of which the most important
class are the single-trace operators of the type: 
\be 
{\cal
  O}_n=\tr\,f_n(F_{\mu\nu},\phi^a,\cdots) 
\ee 
It has been shown\cite{Policastro:2002se} that one can use the AdS/CFT
correspondence to obtain
universal predictions for the transport coefficients of a gauge-theory
plasma, qualitatively (in some ways) similar to the quark-gluon plasma
produced in relativistic heavy-ion collisions.

If we consider boosted extended black holes (really
branes) in AdS space-time having a definite temperature and velocity.
Via holography these get related to the temperature and velocity of
asymptotically AdS solutions to the Einstein equations. 
As a result one finds a
beautiful relation between the Einstein equations and Navier-Stokes
equations, the equations of fluid dynamics (see for example
Refs.\cite{Bhattacharyya:2008jc}).

This approach has practical utility in understanding strongly coupled
plasmas that are hard to study using other more conventional
formalisms. But its most striking feature is that it correlates
fundamental properties of hydrodynamics and gravitation -- a
remarkable synthesis of disparate ideas with a long history.

\subsection{New insights about gravity IV}

The AdS/CFT correspondence tells us that gravity and gauge theory are
not two different types of forces but instead are equivalent to each
other. This may seem very bizarre at first sight. However the
different dimensionalities of the gravity and gauge theory suggest
that the transcription between them is not going to be
straightforward, and locality of the bulk theory is a particularly
striking mystery\footnote{On which steady progress is being made, see
  Ref.\cite{Heemskerk:2009pn}.}. Despite useful examples, the full
dictionary between gravity and gauge theory concepts is complicated
and not fully understood. However one key illumination is that the
radial direction in AdS gravity is related to the energy scale of the
gauge theory. Therefore from the gauge theory point of view, motion
along this direction is seen as a renormalisation group flow!  (see
Ref.\cite{deBoer:1999xf} and for very recent work in this direction,
Ref.\cite{Douglas:2010rc}).

The correspondence described here is a realisation of an older
conjecture known as the ``holographic principle'', which says that the
degrees of freedom of quantum gravity propagating in a bulk region can
be encoded on the boundary of that region. Just as the comparison of
microscopic and macroscopic entropy of black holes in string theory
provided a concrete realisation of something that had previously been
conjectured, here too we have for the first time a concrete
realisation of a principle that had previously been articulated on
physical grounds by 't Hooft\cite{'tHooft:1993gx}, and
Susskind\cite{Susskind:1994vu}.

To date, the AdS/CFT correspondence has primarily been used to gain
information about strongly coupled gauge theories using the dual
classical, weakly-coupled supergravity. This is of physical interest
because of QCD and speculative theories beyond the Standard Model, and
also because of strongly coupled systems in condensed matter theory,
which for reasons of space we have not been able to discuss
here\footnote{But see Ref.\cite{Hartnoll:2009qx} for a review.}.
Information in the reverse direction would be highly desirable in
order to grapple with conceptual problems of quantum gravity. This has
been slightly less forthcoming because the domain in which
gauge theory is relatively easy to study is that of weak coupling, in
which region the dual AdS superstring propagates in a highly curved
space-time of sub-stringy size. This is not directly relevant to the
world today.

However as we briefly discuss below, a sub-stringy holographic world
would be very relevant for studies of the early universe. And of
course the work on black hole entropy described earlier makes use of
information from the gauge-theory side in important ways. Though some
of the initial ideas for deriving microscopic black hole entropy in
string theory were developed before the AdS/CFT correspondence had
been precisely articulated, today these ideas have found a natural
place in the context of holography which has also provided many
generalisations. It is likely that the next decade will see increasing
feedback from the AdS/CFT correspondence and its generalisations to
address conceptual issues in quantum gravity.

\section{Flux compactifications}

The old problem of string theory compactifications, namely the large
number of undetermined moduli or flat directions, also benefitted from
the AdS/CFT correspondence which led to a new line of attack. It had
been known that nontrivial fluxes in the internal manifold could fix
some of the moduli by providing a potential for them. Using
inspiration from AdS spacetime, it was then realised that such fluxes
typically lead to ``warped'' backgrounds of string theory: backgrounds
in which the metric at a given point of the noncompact spacetime
depends on the point in the internal space. Such warped backgrounds
can have an AdS-like ``throat'' region in the internal space and
one can thereby obtain a desirable separation of
scales. Flux compactifications use combinations of D-branes,
anti-branes and orientifolds\footnote{Singularities similar to orbifolds but
which carry RR charge.} to generate, among other things, a
positive cosmological constant and broken
supersymmetry\cite{Kachru:2003aw} (see also earlier work
on flux stabilisation in
Refs.\cite{Strominger:1986uh,Dasgupta:1999ss,Giddings:2001yu}).

Flux compactifications are the closest we have come to realising the
Standard Model of particle physics, as well as a realistic
gravitational sector including a cosmological constant, in the context
of string theory. Their success has paradoxically raised a problem: it
seems likely that string theory admits an immense number of vacua,
known as the ``landscape'', and many of them lie arbitrarily close to
the real world. The problem of finding ``the right vacuum'' in this
situation looks very daunting. We have ended up very far from the
original hope that a simple, almost unique compactification of string
theory would lead to a viable description of the real world.

\section{String Cosmology}

The initial analyses of D-branes focused on charged, stable branes,
and for a while it was not noticed that string theory
admits other D-branes that are uncharged and unstable. One way to
think of them is to consider a pair of a D-brane and an
anti-D-brane, the net charge of this system being zero. Such a system
would generically be unstable and decay.

It has been shown (a review can be found in Ref.\cite{Sen:2004nf})
that the decay of these branes can be understood in terms of a
tachyonic particle that arises from quantising the open string ending
on the brane (this differs from the closed-string tachyon that we
encountered in bosonic string theory). The rolling of the tachyon down
to the bottom of its potential well is the world-volume description of
the process in space-time wherein the brane decays. This ``rolling
tachyon'' paradigm has proved very influential in string-inspired
cosmology.

Cosmology involves the study of regions of space-time (particularly
relevant in the early universe) where conventional physics breaks down
because of the possible occurrence of singularities. Holography can
become a conceptual necessity in situations where the geometric
description of space-time itself is in question -- one may then have
no choice but to shift from a macroscopic picture (gravity) to a dual
microscopic picture. Finally, an ultraviolet completion of general
relativity is required when energies reach the Planck scale.  Because
of the many different properties of string theory described in this
article, it is a prime candidate to address and potentially solve all
these problems (these observations and the ones that follow are
elaborated in the review article Ref.\cite{McAllister:2007bg}).

Discussions about cosmology in string theory are typically made in the
framework of non-supersymmetric configurations of multiple D-branes,
referred to above, with the tachyon playing an important role in the
dynamics. Of course such discussions require a
sensible background space-time, for which one needs at least a
semi-realistic model. One of the most important constraints on such
backgrounds is that the moduli associated to the string
compactification be completely stabilised.

The primary goals of string cosmology would be to provide a precise
mechanism for inflation and an explanation of dark energy -- the
latter being arguably the single biggest puzzle in our microscopic
picture of the world. Importantly, there can be conflicts between such
mechanisms and the mechanism for moduli stabilisation. Also, inflation
can potentially screen from observation the kind of microscopic
information about the early universe that a particular string model
might provide.

In addition to addressing basic open questions about the origin of the
universe, string cosmology potentially provides an experimental
testing ground for string theory itself. This would require
identifying observational features of string theory that are hard, or
impossible, to reproduce in field theory. The nature of corrections to
the cosmic microwave background provides a possible class of examples.
Despite (or perhaps because of) these difficulties, string cosmology
is currently among the most fertile fields of research within string
theory.

\section{Conclusions}

It may not be wrong to say that the relevance of string theory to the
real world it still to be completely determined. However the question
of ``relevance'' in this context has evolved far beyond what anyone could
have imagined in 1985. 

As a candidate for a unified theory of all fundamental interactions,
string theory is extremely compelling but a precise theory with a set
of unambiguous predictions is indeed not yet available. One key
obstacle is the the landscape problem. Far from there being no
accurate string description of the real world, it looks likely that
there could be so many of them as to make the discovery of the
``right'' one virtually impossible.  This is an issue on which new
ideas are awaited.

However, string theory is also a potent and precise formalism that
generalises and probably supplants quantum field theory. It provides
an ultraviolet consistent theory of quantum gravity wherein
singularities are typically smoothed out. At low energies it reduces
to ordinary Einstein gravity and its perturbation expansion can be
used to compute quantum corrections to gravitational processes.
Moreover it has a non-perturbative structure of which many key aspects
are well-understood. This allows it to be used as an unparallelled
and accurate testing ground for concepts in quantum gravity.

String theory has provided the provided the most convincing evidence
that black holes have a large number of microscopic states that
account for their entropy. This has essentially settled in the affirmative
the fundamental question of whether the thermodynamic nature of
gravity arises from microscopic states, just like the relation between
usual thermodynamics and statistical mechanics. It is reasonable to
expect that the information-loss paradox will similarly be addressed,
and perhaps resolved, using string theory in the near future.

The usefulness of strings in providing a holographic description of
the strong force and associated strongly coupled fluids is
demonstrated and steadily growing. It is widely expected that the
actual QCD string, the very basis for the invention of string theory,
will be found in the near future.

The application of string models to study the early universe
promises to uncover the secrets of the big bang. Given the increasing
body of observations pertinent to the early universe, it is
just conceivable that this direction could also yield the long sought-for
experimental tests of string theory.

Perhaps the single most important point is that all the above
applications of string theory do not exist independently of each
other. Gravity, gauge symmetry, compactification, unification,
holography, hydrodynamics, strongly correlated systems and cosmology,
and probably many more aspects of physics, are inextricably linked to
each other in this remarkable and profound edifice.

\bibliographystyle{JHEP}
\bibliography{sm-cqg}

\end{document}